\documentclass[12pt,reqno]{amsart}

\usepackage[left=1in, right=1in, top=1.1in,bottom=1.1in]{geometry}
\usepackage[T1]{fontenc}
\usepackage{amsthm,amssymb,amsmath,amsfonts,mathrsfs,amscd}
\usepackage{enumitem}
\usepackage{stmaryrd}
\usepackage{mathrsfs}

\usepackage{booktabs} 
\usepackage{array}    
\usepackage{makecell} 
\usepackage{multirow}
\usepackage{pdflscape}   
\usepackage{threeparttable}

\usepackage{graphicx}
\graphicspath{ {./fig/}}
\usepackage{float}

\usepackage{pdfsync}
\usepackage[font={scriptsize}]{caption}

\usepackage{hyperref}
\hypersetup{
  colorlinks   = true,
  citecolor    = green,
  linkcolor    = red
}

\theoremstyle{plain}

\theoremstyle{definition}

\theoremstyle{remark}

\begin{document}


\title{Stylized facts in Web3}
\author[A. C. Silva]{A. Christian Silva}
\author[S.-N. Tung]{Shen-Ning Tung}
\author[W.-R. Chen]{Wei-Ru Chen}

\address{A. Christian Silva \newline
IdataFactory, Stuart, FL, USA
}
\email{csilva@idatafactory.com}

\address{Shen-Ning Tung \newline
Department of Mathematics, \newline
National Tsing Hua University \newline
Hsinchu, Taiwan
}
\email{tung@math.nthu.edu.tw}

\address{Wei-Ru Chen \newline
Department of Mathematics, \newline
National Tsing Hua University \newline
Hsinchu, Taiwan
}
\email{std92050@gmail.com}



\begin{abstract}
This paper presents a comprehensive statistical analysis of the Web3 ecosystem, comparing various Web3 tokens with traditional financial assets across multiple time scales. We examine probability distributions, tail behaviors, and other key stylized facts of the returns for a diverse range of tokens, including decentralized exchanges, liquidity pools, and centralized exchanges. Despite functional differences, most tokens exhibit well-established empirical facts, including unconditional probability density of returns with heavy tails gradually becoming Gaussian and volatility clustering. Furthermore, we compare assets traded on centralized (CEX) and decentralized (DEX) exchanges, finding that DEXs exhibit similar stylized facts despite different trading mechanisms and often divergent long-term performance. We propose that this similarity is attributable to arbitrageurs striving to maintain similar centralized and decentralized prices. Our study contributes to a better understanding of the dynamics of Web3 tokens and the relationship between CEX and DEX markets, with important implications for risk management, pricing models, and portfolio construction in the rapidly evolving DeFi landscape. These results add to the growing body of literature on cryptocurrency markets and provide insights that can guide the development of more accurate models for DeFi markets.
\end{abstract}


\maketitle

\section{Introduction}
The financial landscape has undergone a significant transformation with cryptocurrencies, spearheaded by Bitcoin (BTC) \cite{nakamoto2008bitcoin}. The growing integration of cryptocurrencies into investor portfolios and the financial industry's increasing interest, evidenced by recently created Bitcoin-based ETFs, underscore their rising prominence. However, Bitcoin represents merely the tip of the iceberg in digital finance.

The introduction of the Ethereum network \cite{EthereumWhitepaper} marked a pivotal moment, enabling the creation of an ecosystem of decentralized applications now collectively known as Web3, with Decentralized Finance (DeFi) as its most successful example \cite{harvey2021defi}. This development has expanded the cryptocurrency offering beyond simple Bitcoin clones to essential components of a new financial paradigm.

In this evolving landscape, the term "cryptocurrency" has become limiting. "Tokens" more accurately describe the diverse array of digital assets. For instance, Ether (ETH) serves as the native token of the Ethereum network, facilitating smart contract execution. Similarly, liquidity provider tokens like CRV allow holders to earn fees for liquidity on decentralized exchanges such as Curve while also being tradable assets. The rapid expansion of DeFi over the past five years has given rise to applications addressing a broad spectrum of financial needs, as detailed in \cite{harvey2021defi}.

A unique feature of Web3 is that every token can be traded and has a price. This stems from its inherent decentralization, which leads to function fragmentation. Consequently, every project tends to have its token, irrespective of whether it was created as a medium of exchange like Bitcoin.

This paper presents a comprehensive statistical analysis of various tokens within the Web3 ecosystem. We report a series of stylized facts and contrast these with well-documented stylized facts in traditional equity markets \cite{Mantegna_Stanley_1999, Cont2001, Bouchaud_Potters_2003, Chakraborti2011}. Our analysis aims to characterize the main tokens in circulation thoroughly, potentially guiding future models and contributing to Web3's further development.

Moreover, we compare centralized exchanges (CEX) with decentralized exchanges (DEX), offering insights into the evolving dynamics between traditional and decentralized financial infrastructures.

Our study examines several key stylized facts, including fat-tailed return distributions, aggregation normality, autocorrelation properties, volatility clustering, leverage effects, time-reversal asymmetry, and factors driving cross-sectional returns.

\section{Blockchain and Tokens in Web3 Ecosystems}

\subsection{Blockchain Technology}

Blockchain technology is the cornerstone of Web3 systems, comprising a chain of information blocks interconnected and secured through cryptographic keys \cite{lipton2022blockchain}. This distributed ledger technology ensures decentralized storage and transparency across a global network of computers. While initially serving primarily as a transaction record for Bitcoin \cite{nakamoto2008bitcoin}, blockchain technology has evolved to support more complex functionalities, particularly with platforms like Ethereum.

A primary challenge in decentralized systems is ensuring transaction uniqueness and preventing double-spending. Consensus mechanisms such as "proof-of-work" (PoW) or "proof-of-stake" (PoS) \cite{seang2019proof} address this by requiring network participants (miners or validators) to validate and agree on the ledger's state.

\subsection{The Role of Tokens in Web3}

Tokens play a crucial role in blockchain ecosystems, serving various functions:

\begin{enumerate}
    \item L1 Tokens: Native to independent blockchains (e.g., BTC, ETH)
    \item L2 Tokens: Associated with scaling solutions built on L1 networks (e.g., ARB, OP)
    \item DeFi Tokens: Used in decentralized finance applications (e.g., UNI, CRV)
    \item GameFi Tokens: Utilized in blockchain-based games (e.g., AXS, SAND)
    \item Meme Coins: Often created as social experiments or speculative assets (e.g., DOGE, SHIB)
    \item Stablecoins: Designed to track a given fiat currency such as the USD.
\end{enumerate}

The power conferred by different tokens varies widely. For instance, native tokens like ETH enable network validation and smart contract execution, while governance tokens offer voting rights in protocol decisions (similar to traditional stocks).

\subsection{Tokens in Our Analysis}
\begin{figure}
\centering
\includegraphics[width=1\linewidth]{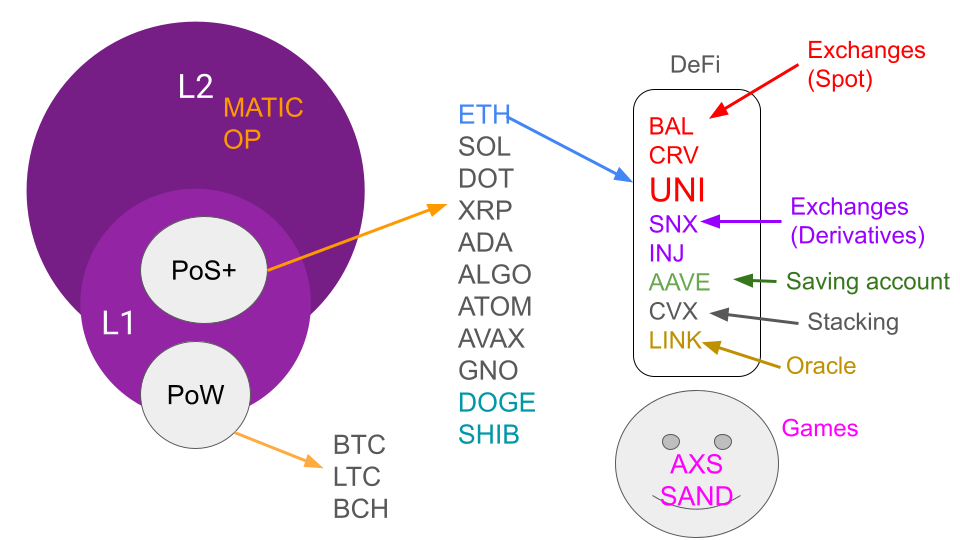}
\caption{List of tokens studied, classified by application and consensus mechanism}
\label{fig:tok}
\end{figure}

Our analysis encompasses various tokens representing the Web3 universe, as Figure \ref{fig:tok} illustrates. This includes multiple L1 chains utilizing different consensus mechanisms (PoW vs PoS), L2 solutions, and tokens built on the Ethereum network. We particularly emphasize UNI, the native token of Uniswap, a prominent decentralized exchange.

In addition to the many tokens, we will contrast decentralized and centralized exchanges using ETH, which are traded on both. 

\subsection{Centralized vs Decentralized Exchanges}

Exchanges are fundamental in Web3 because every token, regardless of its role in a given decentralized application, is also a tradeable asset. This ability to trade is essential because a user paid on a given token might need to purchase something else with another token. Therefore, we study whether a given exchange will affect the statistical features of a traded token.

When it comes to Web3, there are several exchange choices. A key distinction in the cryptocurrency ecosystem is between Centralized Exchanges (CEX) and Decentralized Exchanges (DEX):

\begin{itemize}
    \item CEXs (e.g., Binance, Coinbase): Operate similarly to traditional stock exchanges, with a central authority managing order books and asset custody. They serve as main entry points for fiat-to-crypto transactions.
    \item DEXs (e.g., Uniswap, Curve): Operate on blockchain networks using smart contracts for peer-to-peer trading. They use liquidity pools and fixed algorithms instead of order books, offering greater transparency and reduced counterparty risk but potentially lower liquidity and higher transaction costs.
\end{itemize}

CEXs are very well known and account for the most significant trading volumes. Binance alone trades around \$22 billion daily on the spot market and another \$68 billion in derivatives, according to coinMarketcap. DEXs on the other hand, only represent about 10\% of CEX's volume. However, DEXs are fundamental for a fully on-chain application stack. 

It is important, therefore, to note that if one wants to buy a given token, say ETH, one can not just use regular fiat currency such as the USD. That can be done at a CEX but not at a DEX. The closest one can do is first use a CEX to convert USD to a stablecoin such as USDC, USDT, or DAI and then exchange the stablecoin for ETH on a DEX (L1 or L2 network). Therefore, when looking at stylized facts on a DEX, we will look at the pool that exchanges stablecoins into ETH on Mainnet (L1) or Arbitrum (L2). We will compare these pools' prices to a CEX ETH price.\footnote{We distinguish pools by stablecoin and also by whether they utilize the main Ethereum L1 blockchain (called "mainnet") or an alternative L2 network, in our case Arbitrum (Arb). Arbitrum is one of the many scaling solutions to the Ethereum network, which speeds up transaction times and, therefore, reduces trading fees.}

\section{Data}
This section outlines the data sources, selection criteria, and dataset characteristics employed in our study of Web3 tokens and related financial instruments.

\subsection{Data Sources and Selection Criteria}
Our study utilizes a diverse dataset encompassing decentralized finance (DeFi) and traditional financial markets. To ensure data reliability and relevance, we applied strict selection criteria. We focused on tokens associated with teams or protocols operational with established user bases for at least two years. Additionally, we excluded stablecoins due to their primary liquidity being on decentralized exchanges (DEXs), which results in reported prices relative to each other without a reliable baseline for comparison.

Our data sources include Centralized Exchanges (CEX), specifically token prices from Binance and Coinbase \footnote{Code to download data: \url{https://github.com/silvaac/token_style}}, two of the largest and most liquid cryptocurrency exchanges. We also gathered data from Decentralized Exchanges (DEX), focusing on Ethereum (ETH) prices from Uniswap, a leading DEX. Here, we considered 5 basis point (bp) and 30 bp liquidity pools (L1 and L2), examining ETH pairs with major stablecoins: USDC, USDT, and DAI. For comparison with traditional financial markets, we included SPY ETF prices as a proxy for the S\&P 500 index and the EUR to USD exchange rate.

\subsection{Data Collection Challenges and Methodological Implications}
A notable aspect of our data collection process is the inherent transparency and accessibility of blockchain-based financial data. Unlike traditional finance, where data access often comes at a significant cost, Web3 data is public by design. This openness facilitates our research and highlights a fundamental difference between traditional and decentralized finance ecosystems.

However, the relatively short history of many tokens (often just a few years) presents challenges for long-term analyses. To address this limitation, we focused on higher-frequency data, specifically hourly and daily returns, to ensure robust statistical analyses. This approach allows us to extract meaningful insights despite the limited historical depth.

\subsection{What Do We Model ?}
Following standard financial literature practice, we analyze stylized facts using log-returns [$\log{(\text{price}(T))}-\log{(\text{price}(T-t))}$] over a given time interval $t$.In this work, $t$ is, for the most part, one hour, but we will also look at $t$ on the order of days and months when illustrating the central limit theorem and shorter (minutes) when looking at DEX price formation. 

We choose log-returns because it allows us to benchmark our results against traditional finance assets in the literature, which normally use log-returns. However, it's important to note that the choice of log-returns has potential drawbacks, particularly for Web3 tokens, given their large volatility. As discussed in \cite{Bouchaud_Potters_2003}, log-returns can potentially create spurious effects and represent a mathematical abstraction that cannot be directly traded.

In what follows, we call "log-returns" simply as "returns" unless otherwise noted.

\section{Stylized Facts}
Stylized facts are well-established empirical observations consistently found in financial market data \cite{Mantegna_Stanley_1999, Cont2001, Bouchaud_Potters_2003, Chakraborti2011}. They provide a qualitative framework of accepted characteristics for asset and market models, emphasizing the importance of detailed empirical analysis before proposing theoretical models. This approach, akin to experimental sciences like physics, has been fundamental in developing econophysics \cite{Mantegna_Stanley_1999}.

While several studies have documented stylized facts for cryptocurrencies, particularly Bitcoin and Ethereum (e.g. \cite{ZHANG2019598, 2024arXiv240211930T}), our study expands this analysis to a broader, curated list of tokens representing various Web3 applications (Figure \ref{fig:tok}). Notably, we compare Ethereum (ETH) traded on both decentralized (DEX) and centralized exchanges (CEX), offering novel insights into the evolving crypto market structure.

Although the list of stylized facts can be extensive, we focus on seven key phenomena widely observed in equity markets:

\begin{enumerate}
    \item Fat-tailed unconditional probability distribution for price returns over time intervals
    \item Aggregation normality: unconditional probability distribution for price returns becomes normal for very long time intervals
    \item Absence of auto-correlation for price returns, except potentially for short periods
    \item Volatility clustering
    \item Leverage effect
    \item Time-reversal asymmetry
    \item Few factors driving the cross-section of returns
\end{enumerate}

\subsection{Fat-tails and Aggregated Normality}
One of the most unequivocal stylized facts in finance is the presence of fat-tails in the unconditional probability distribution of returns for time intervals typically less than a few days. This phenomenon implies that rare events are more likely to occur than what a Gaussian distribution would predict, hence the terms "fat-tails" or "heavy-tails" \cite{Chakraborti2011}. It has also been observed that this probability distribution gradually approaches a Gaussian distribution (i.e., tails become less fat) for returns over longer periods, typically weeks or months. We refer to this convergence towards a Gaussian distribution as aggregated normality. In this section, we document these well-established stylized facts for Web3 tokens.

\subsubsection{Probability Density Function Analysis}
\begin{figure}[ht]
\centering
\includegraphics[width=1\linewidth]{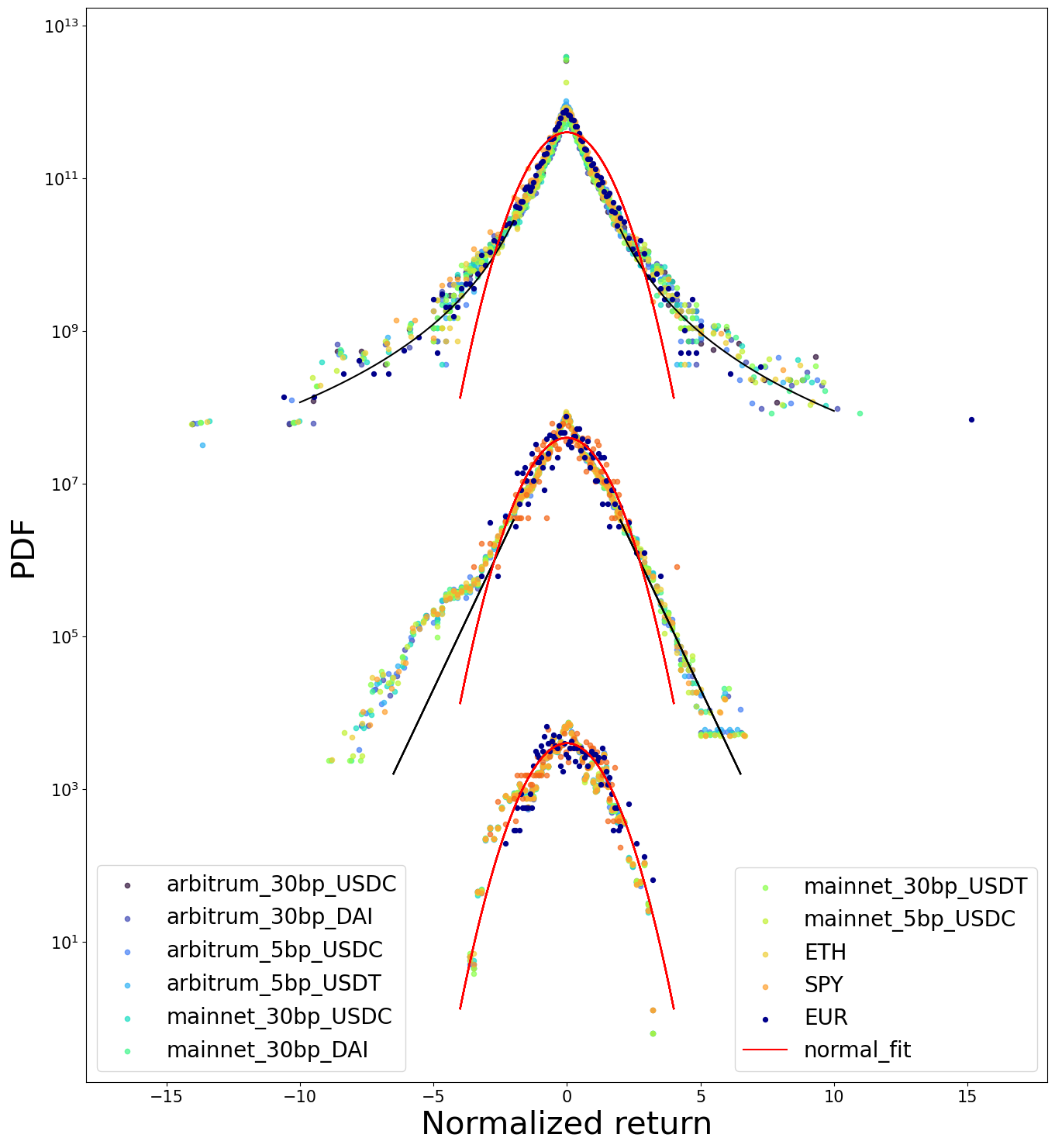}
\caption{PDF for pool, CEX (Binance) ETH, and traditional finance returns at one hour (1H), one day (1D), and one month (1M) shown in a log-linear scale. PDFs are shifted for visualization: multiplied by $10^{12}$ for 1H, $10^8$ for 1D, and $10^4$ for 1M. Assets include various Arbitrum and Mainnet pools, Binance ETH, SPY, and EUR. Power law fit for 1H: $\sim \lvert x \rvert^{-\eta}$ for $\lvert x \rvert \geq 2$ where $\eta = 3.4$; exponential fit for 1D: $\sim e^{-\eta \lvert x \rvert}$ where $\eta = 1.7$.}
\label{fig:pool_CEX_pdf}
\end{figure}

Figure \ref{fig:pool_CEX_pdf} presents the empirical unconditional probability density function (PDF) of normalized returns over one hour, one day, and one month for ETH traded on CEX and DEX, as well as traditional assets such as SPY and EUR.

The figure clearly shows that one-hour returns exhibit fat-tails which are well fit to a power-law ($\sim |x|^{-\alpha-1}$) with $\alpha \sim 2.4$, in agreement with the financial literature \cite{Chakraborti2011}. The one-day returns are closer to a standard normal at the center, and the tails have become asymmetric and closer to exponential ($\sim \exp{(-\eta |x|)}$) with $\eta \sim 1.7$, close to the $\sqrt{2}$ predicted by \cite{SILVA2004227} for stocks. Finally, at one month, the returns are well approximated by a standard normal probability distribution (red line).

\begin{figure}[ht]
\centering
\includegraphics[width=1\linewidth]{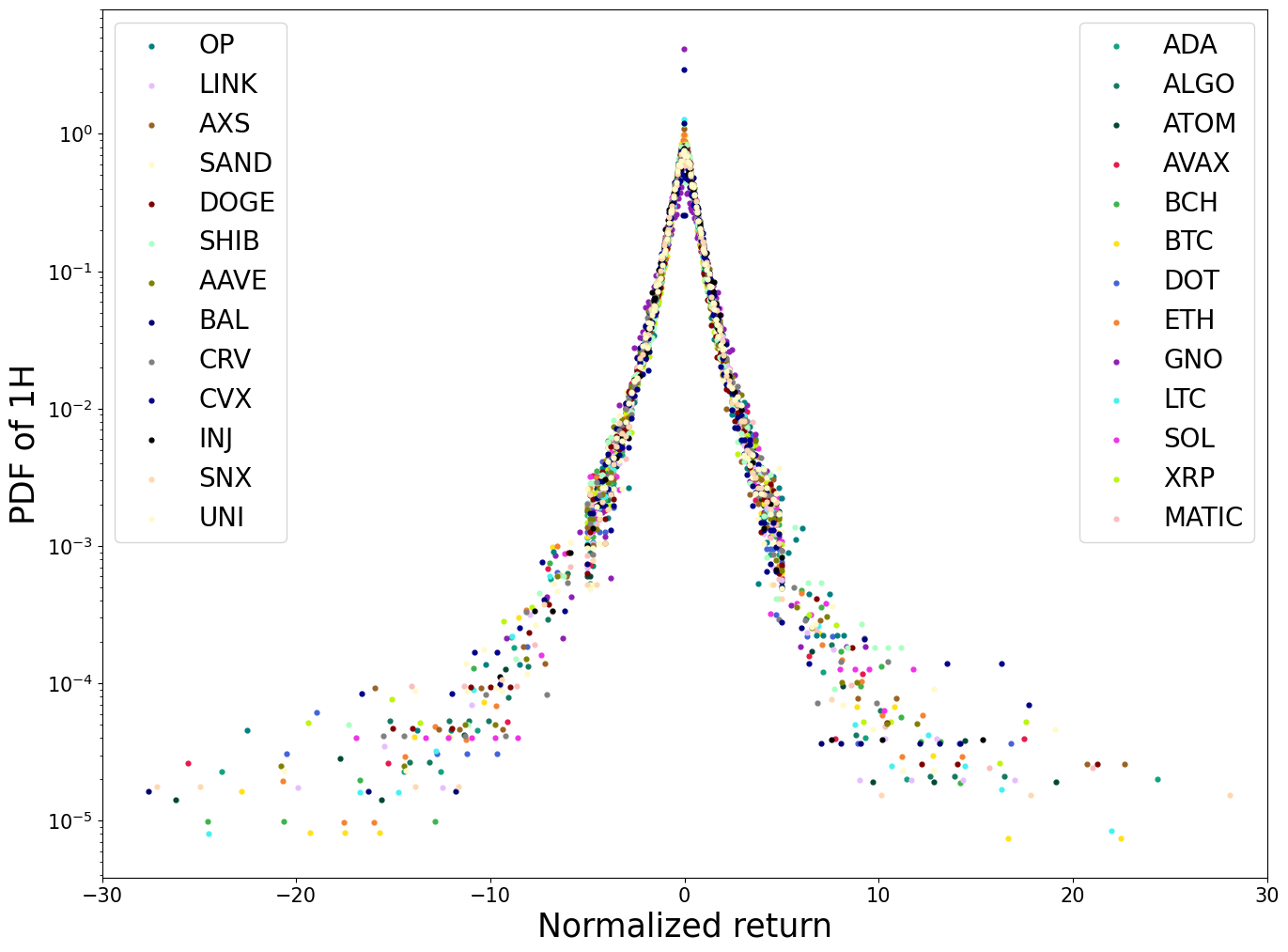}
\caption{Probability density function (PDF) of one-hour normalized returns for various tokens traded on Coinbase. Note: XRP data is sourced from Binance due to its temporary delisting from Coinbase between early 2021 and late 2023.}
\label{fig:coinbase_pdf}
\end{figure}

Figure \ref{fig:coinbase_pdf} shows the one-hour returns PDF for all Tokens in Table \ref{tab:asset_table}. The tails are power law in agreement with ETH in Figure \ref{fig:pool_CEX_pdf} and with the fat-tail stylized fact.

In Appendix \ref{app:tail}, we analyze the cumulative distribution function (CDF) and fit the tails of one-hour returns to a power law ($\text{CDF}(x) \sim |x|^{-\alpha}$) for returns exceeding two standard deviations. We find that the power coefficient $\alpha > 2$, typically between 2 and 3 with an average of 2.6 (Table \ref{tab:asset_table}), aligns with traditional finance observations \cite{Chakraborti2011}. 

\subsubsection{Jarque-Bera Test for Normality}
To further quantify the convergence to the normal distribution, We employ the Jarque-Bera (JB) test statistic, which detects deviations from normality due to skewness or excess kurtosis. The JB test statistic for a random variable $X$ is defined as:
\begin{equation}
\text{JB}(X) = \frac{n}{6}\left(s^2(X) + \frac{1}{4}(k(X) - 3)^2\right)
\end{equation}

\noindent where $n$ is the number of observations, $s(X)$ is the sample skewness, and $k(X)$ is the sample kurtosis. Under the null hypothesis of normality, the JB statistic asymptotically follows a chi-squared distribution with two degrees of freedom.

\begin{figure}[ht]
\centering
\includegraphics[width=1\linewidth]{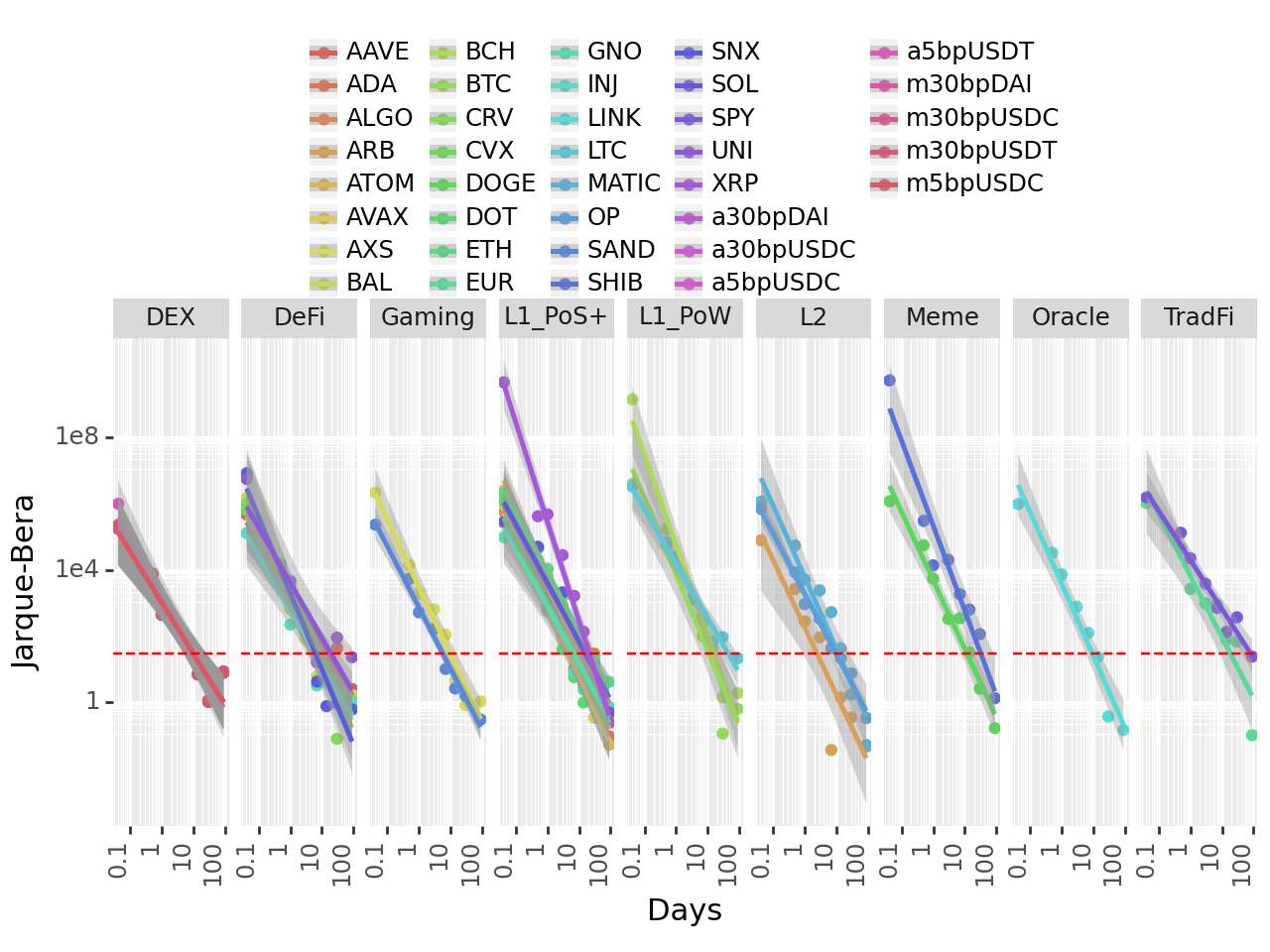}
\caption{Jarque-Bera statistics for tokens, SPY, and EUR against return in days, with linear fit including error bars (gray shaded area). The red line indicates the 95\% confidence level for Gaussian distribution.}
\label{fig:pool_CEX_JB_stats}
\end{figure}

Figure \ref{fig:pool_CEX_JB_stats} displays the JB statistic for all studied tokens (Figure \ref{fig:tok} and Table \ref{tab:asset_table}) across different return intervals, along with a linear fit to the log-log plot for each asset. Lower JB values indicate closer adherence to a normal distribution.

Analysis of Figure \ref{fig:pool_CEX_JB_stats} reveals several key insights. We cannot reject the null hypothesis of zero skew and zero excess kurtosis with 95\% confidence for most tokens for returns larger than a few days. This finding suggests that the return distributions of most tokens approach normality over longer time horizons. However, exceptions exist: LTC and SPY require returns larger than one month to approach normality, indicating that these assets may have more persistent non-Gaussian characteristics.

The speed of convergence towards normality varies among assets, as evidenced by the slopes of the linear fits in the log-log plot. These slopes range from -1.5 for SPY and ETH on DEX, indicating a slower convergence, to -3 for XRP, suggesting a faster approach to normality. The majority of assets, however, cluster around a slope of -2, as shown in Table \ref{tab:asset_table}, column "JB". This variation in convergence speeds provides insight into the different dynamics governing the return distributions of various assets in traditional and decentralized finance markets.

\subsection{Autocorrelation of Returns}
This section examines the autocorrelation function (ACF) of one-hour returns for various financial assets, including Web3 tokens, SPY, and EUR. In traditional finance, the ACF for returns is generally reported to be insignificant except for short intervals, supporting the efficient market hypothesis as a reasonable approximation.

Our analysis accounts for the different trading schedules of these assets. While tokens trade continuously, SPY trades only from 9:30 to 16:00 Eastern time on weekdays, excluding holidays. We calculate the autocorrelation for all assets over the same historical period, with details on our methodology for handling different exchange sessions provided in Appendix \ref{app:acfR}.

To standardize our analysis across assets with varying data lengths, we scale the ACF values by multiplying them by the square root of the number of data points. This allows us to present consistent error bars at $\pm 3$, corresponding to a 99\% confidence interval, regardless of each asset's data length.

\begin{figure}[ht]
\centering
\includegraphics[width=1\linewidth]{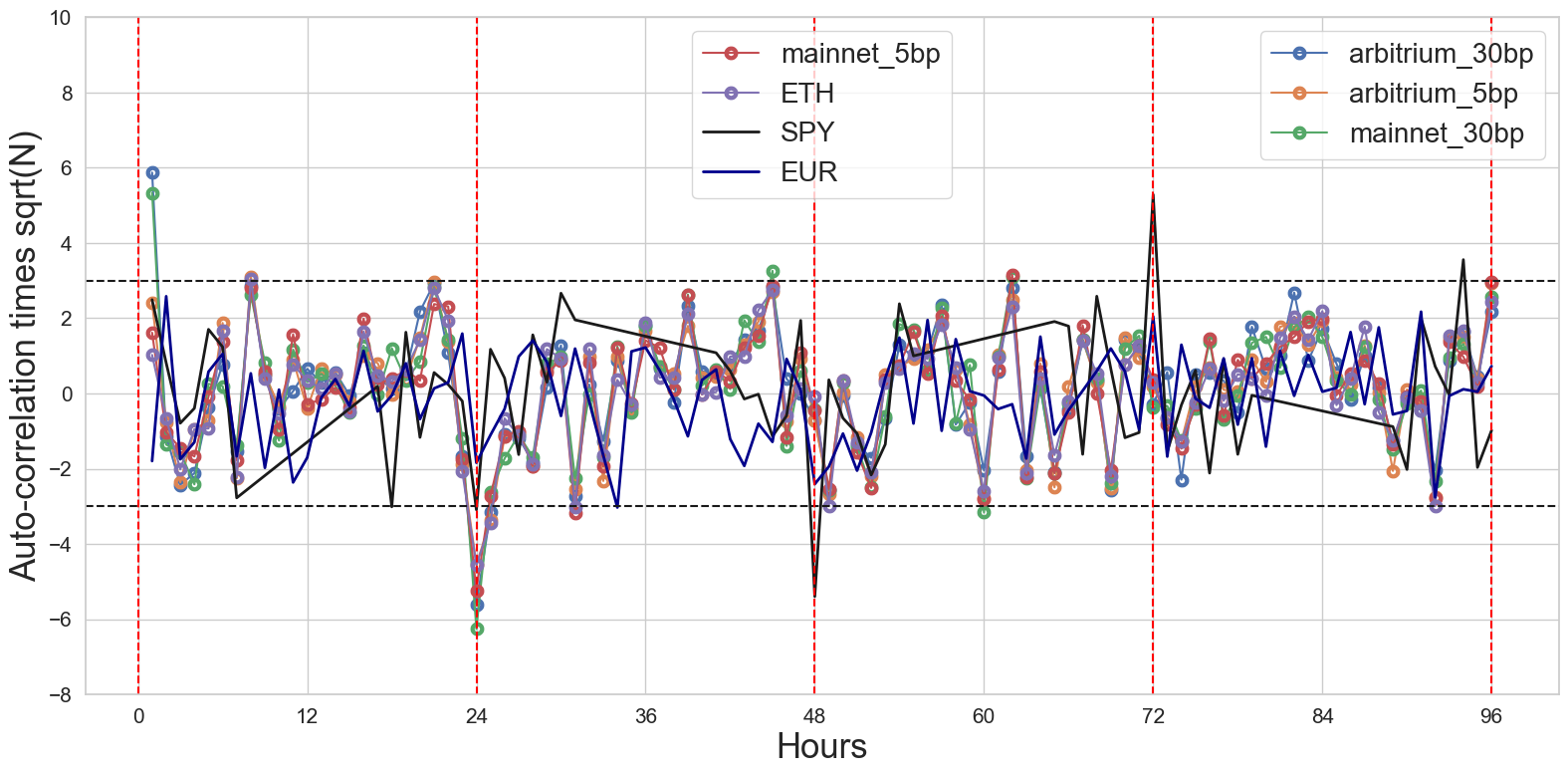}
\caption{Autocorrelation of pool, CEX, and traditional finance assets. Tokens include Arbitrum 30bp WETH \& USDC, Arbitrum 5bp WETH \& USDC, Mainnet 30bp WETH \& USDC, Mainnet 5bp WETH \& USDC, Binance ETH.}
\label{fig:Pool_CEX_trad_acf_R}
\end{figure}

Figure \ref{fig:Pool_CEX_trad_acf_R} illustrates the ACF of returns for ETH on both centralized (CEX) and decentralized exchanges (DEX), alongside SPY and EUR. Results for other tokens, which are qualitatively similar, can be found in Appendix \ref{app:acfR}.

Analysis of Figure \ref{fig:Pool_CEX_trad_acf_R} reveals that while most lags in the ACF are insignificant, validating the stylized fact for our tokens, significant autocorrelations are observed within the first hour and at 24 hours. Notably, Mainnet and Arbitrum Uniswap trading pools (30bp) show significant positive ACF values in the first hour, whereas Binance ETH aligns with SPY and EUR, showing insignificant first-hour ACF. A significant negative autocorrelation is observed at 24 hours, repeating less pronouncedly at 48 hours, echoing findings by \cite{HESTON2008418}. These observations suggest distinct behaviors between DEX and CEX environments, as well as potential cyclical patterns in crypto trading.

\begin{figure}[ht]
\centering
\includegraphics[width=1\linewidth]{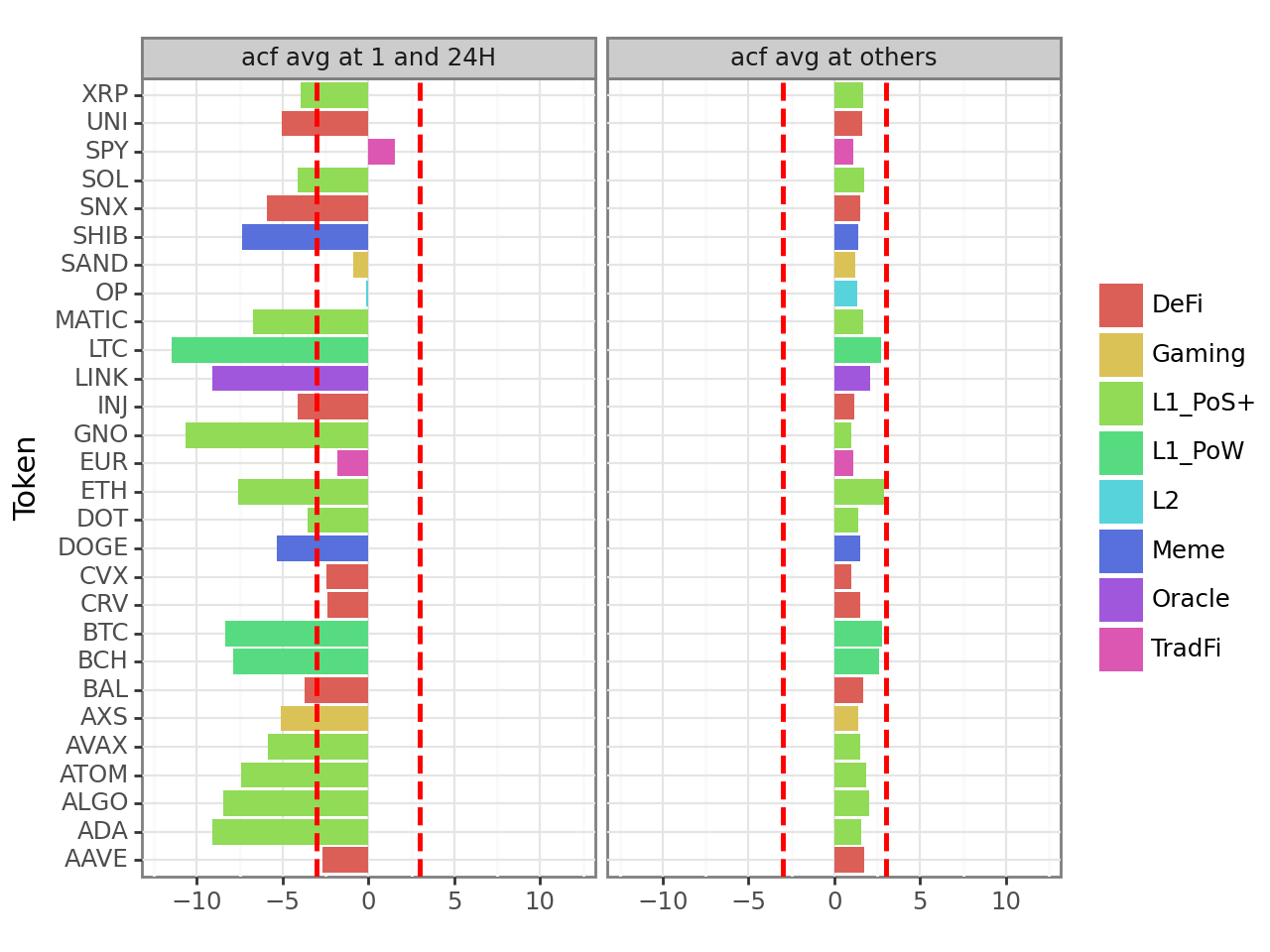}
\caption{Average autocorrelation of the return multiplied by the square root of the number of hours with error bars (red dashed lines). Left: average of lag one and lag 24. Right: average of all 96 lags excluding lag one and lag 24. Color code identifies each token sector.}
\label{fig:acf_sum}
\end{figure}

Figure \ref{fig:acf_sum} provides a comprehensive summary of autocorrelation results across all tokens, utilizing their full history.

The left panel of Figure \ref{fig:acf_sum} highlights the distinct, often negative, and significantly non-zero average autocorrelations for the first and twenty-fourth hours, underscoring the importance of these specific lags in short-term price dynamics. In contrast, the right panel, displaying average autocorrelations for all other lags, shows insignificant values. This corroborates the general lack of substantial autocorrelation, aligning with the efficient market hypothesis beyond very short-term effects.

\subsection{Volatility Clustering}
Volatility clustering, a well-documented stylized fact in financial data, is characterized by the tendency of high (low) volatility periods to follow similar high (low) volatility periods. This phenomenon manifests in return time series as patches or clusters of returns with similar magnitudes.

To quantify this effect, we analyze volatility's autocorrelation function (ACF), which typically exhibits high persistence over extended periods, potentially lasting weeks. We employ the absolute value of hourly returns as a proxy for volatility, noting that our conclusions remain robust to the choice of proxy.

For consistency with our previous analyses and to account for different exchange sessions, we apply a similar algorithm to calculate the autocorrelation of absolute returns for traditional finance assets (SPY, EUR) as used for the autocorrelation of returns. The detailed methodology is available in Appendix \ref{app:acfR}.

Figure \ref{fig:Pool_CEX_trad_acf_absR} presents the ACF of the absolute value of one-hour returns for ETH traded on various exchanges, alongside SPY and EUR, over the same historical period.

\begin{figure}[ht]
\centering
\includegraphics[width=1\linewidth]{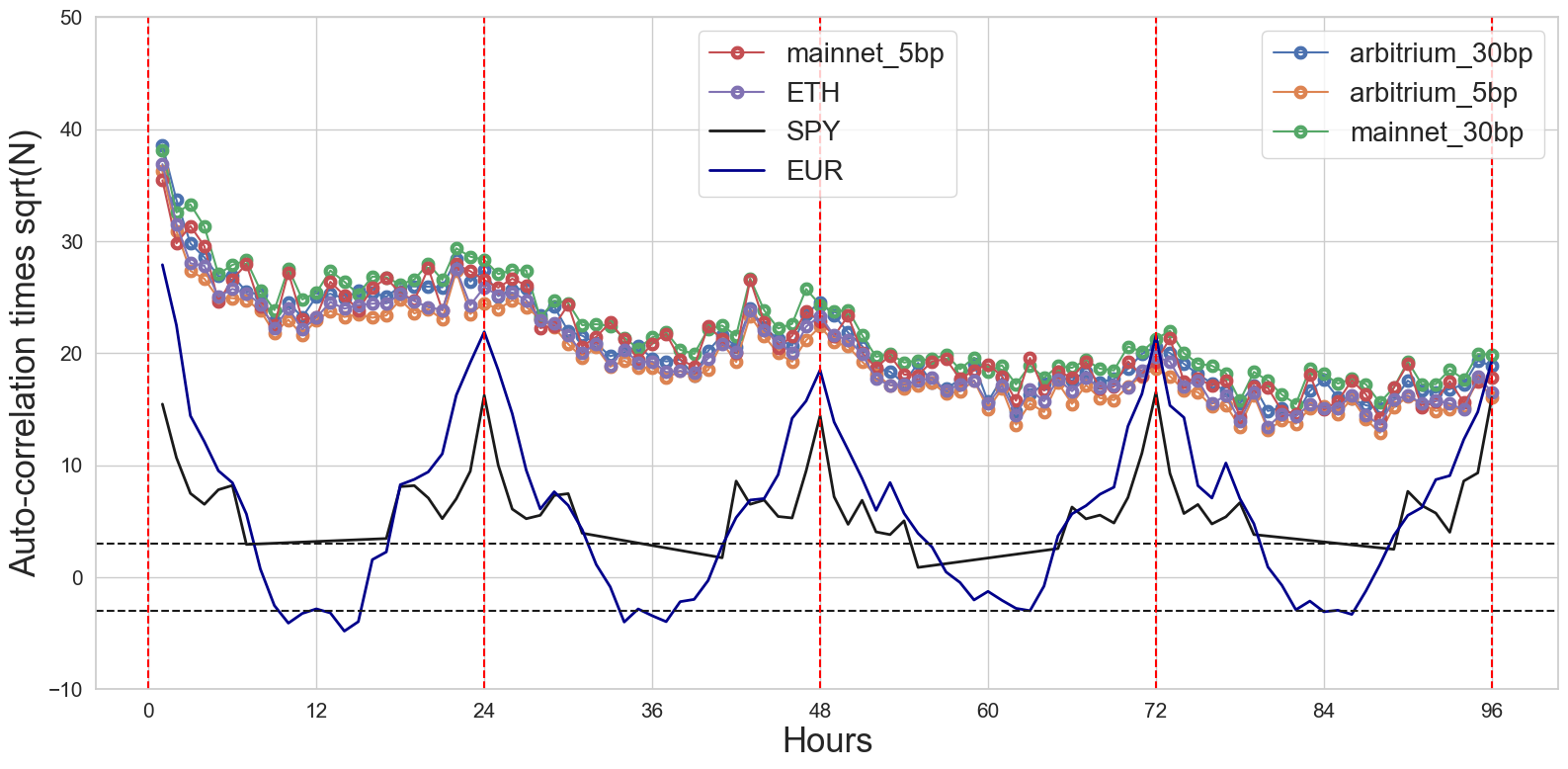}
\caption{ACF of DEX pools, CEX, and traditional finance for one-hour absolute returns over the same historical period. Results show substantial and persistent autocorrelation that is larger than in traditional markets.}
\label{fig:Pool_CEX_trad_acf_absR}
\end{figure}

Analysis of Figure \ref{fig:Pool_CEX_trad_acf_absR} reveals several key insights. The ACF for ETH is unequivocally significant, with all curves substantially above the $\pm 3$ (99\%) error bar. In contrast, SPY and EUR exhibit lower ACF with substantial seasonality due to session closures. Notably, ETH displays a subtle seasonality despite continuous trading, suggesting uneven daily trading activity.

These results are qualitatively consistent across all tokens in Table \ref{tab:asset_table}. Appendix \ref{app:acfR} provides additional ACF plots for various tokens, utilizing all available historical data for each token.

In line with traditional finance, all studied tokens exhibit slow decay in the absolute return ACF, indicative of volatility clustering. The persistence and magnitude of ACF generally exceed those of SPY, even when accounting for differences in data length. Some tokens, such as GNO and CVX, show initial ACF levels comparable to SPY, potentially due to a substantial number of zero one-hour returns (see Table \ref{tab:asset_table}). Appendix \ref{app:zero} offers a simulation illustrating the impact of zero returns on the ACF.

To quantify volatility clustering, we conduct a linear regression of the log of the number of hours (x-axis in Figure \ref{fig:Pool_CEX_trad_acf_absR}) against the log of the ACF of the absolute returns. Table \ref{tab:asset_table} presents the resulting slope and intercept values.

According to \cite{Chakraborti2011}, typical power-law exponents (slope of a log-log fit) for stocks range between -0.1 and -0.4. Our analysis yields values between -0.15 for AAVE and -0.35 for GNO or -0.34 for SHIB. These values fall well within the expected range for stock markets, further corroborating the stylized fact of volatility clustering in crypto assets.

\subsection{Leverage Effect}
The leverage effect, a well-documented phenomenon in equity markets, is characterized by a negative correlation between asset returns and volatility, particularly over short time horizons. This effect suggests that negative returns are typically associated with higher subsequent volatility, while positive returns are associated with lower volatility. Notably, the leverage effect is primarily observed in equities and is not typically seen in other asset classes, such as currencies.

To quantify the leverage effect, we employ a correlation measure between returns and absolute returns, defined as:
\begin{equation}\label{eq:lev}
L(k) \equiv \frac{\mathbb{E}\left[ (\lvert R_{t} \rvert -\mu_{\lvert R \rvert}) (R_{t+k} - \mu_R)\right]}{\sigma_{\lvert R \rvert} \sigma_R}
\end{equation}

\noindent where $R_t$ represents the return at time $t$, $|R_t|$ is the absolute return (proxy for volatility), $\mu_R$ and $\mu_{|R|}$ are the mean values of returns and absolute returns, respectively, $\sigma_R$ and $\sigma_{|R|}$ are the standard deviations of returns and absolute returns and $k$ represents the time lag.

\begin{figure}[ht]
\centering
\includegraphics[width=1\linewidth]{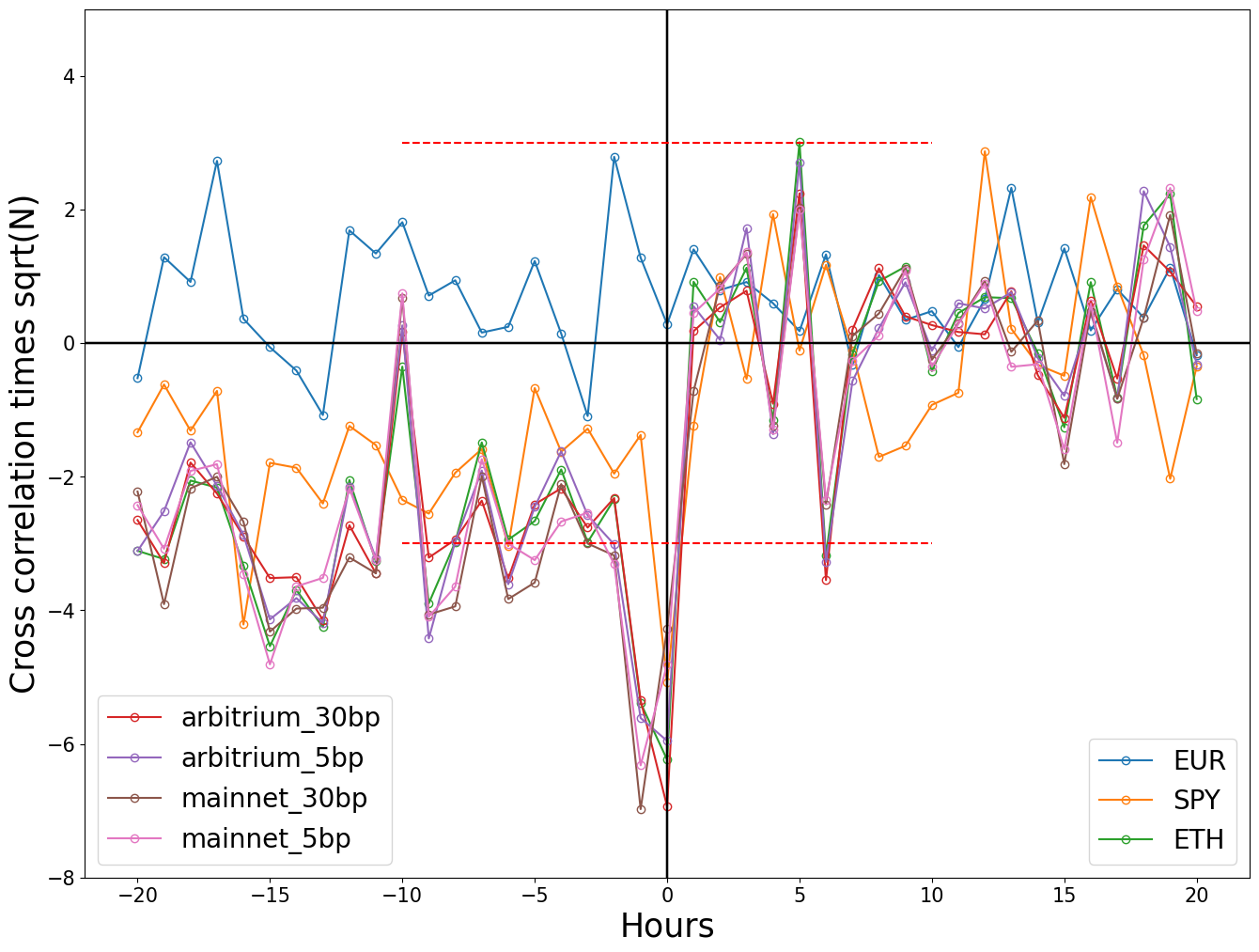}
\caption{Cross Correlation (Equation~\eqref{eq:lev}) between absolute return and return where these values are multiplied by the square root of the number of hours. Red dashed horizontal lines show $\pm 3$ SE (99\%) error bars.}
\label{fig:pool_CEX_lev}
\end{figure}

Figure \ref{fig:pool_CEX_lev} illustrates the correlation (Equation~\eqref{eq:lev}) between past returns and future absolute returns on the negative axis and the correlation between past absolute returns and future returns on the positive axis for ETH (CEX and DEX) and traditional assets over the same historical period. The correlation is multiplied by the square root of the number of hours, consistent with other correlation figures in this study.

Our analysis reveals that the leverage effect is less pronounced than other stylized facts. While most cross-correlation lags in Figure \ref{fig:pool_CEX_lev} are insignificant ($-3<L(k)<3$), there is a clear visual distinction between negative ($k<0$) and positive ($k>0$) lags. The negative lags show a cross-correlation curve oscillating around $L(k)=-3$ (significance level) with significant values ($L(k)<-3$) close to zero hour/lag. SPY exhibits less significant but persistently negative values for hours less than zero, while the curves for positive hours/lags oscillate around zero and are well within the error bars. EUR shows an oscillation of about zero correlation for both positive and negative lags.

It is well established that stock indices exhibit the leverage effect \cite{Bouchaud_Potters_2003}, while currencies do not \cite{Z2009}. Based on these literature findings and our results in Figure \ref{fig:pool_CEX_lev}, we can confidently state that ETH shows the leverage effect, albeit with lower significance than other stylized facts in this study. However, we are less certain about generalizing the leverage effects for other tokens.

\begin{figure}[ht]
\centering
\includegraphics[width=1\linewidth]{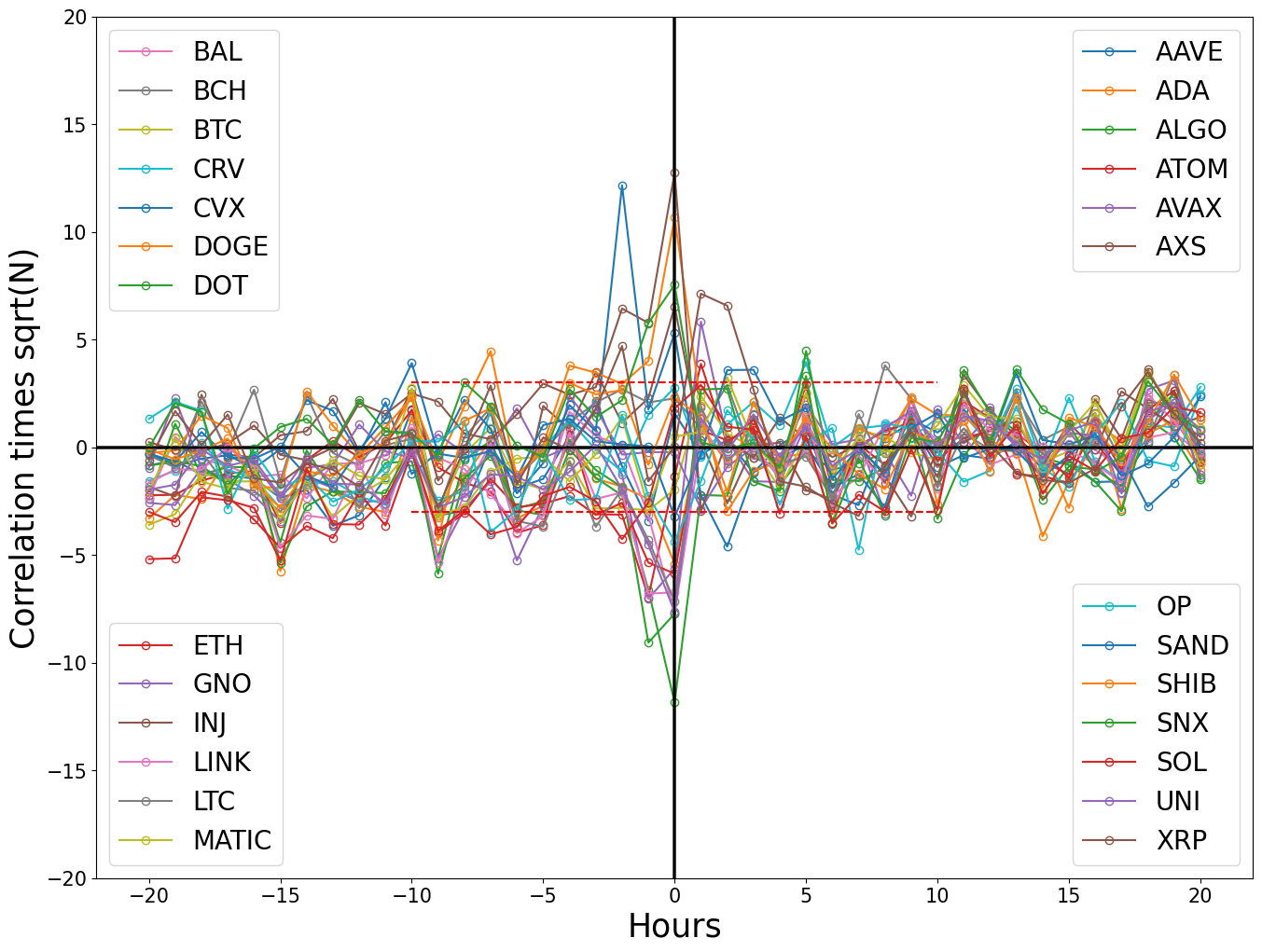}
\caption{Cross-correlation (Equation~\eqref{eq:lev}) for all Coinbase tokens in Table \ref{tab:asset_table} with error bars (red dashed horizontal lines). Most $L(k)$ values are within the error bars for $k>0$.}
\label{fig:cb_lev}
\end{figure}

Figure \ref{fig:cb_lev} presents the $L(k)$ plot for all tokens in Table \ref{tab:asset_table}. The negative ($k<0$) lags do not show a consistent pattern across all tokens, while the positive ($k>0$) lags appear more synchronized for $k>5$ but with substantial differences between tokens when $k<5$.

\begin{figure}[ht]
\centering
\includegraphics[width=1\linewidth]{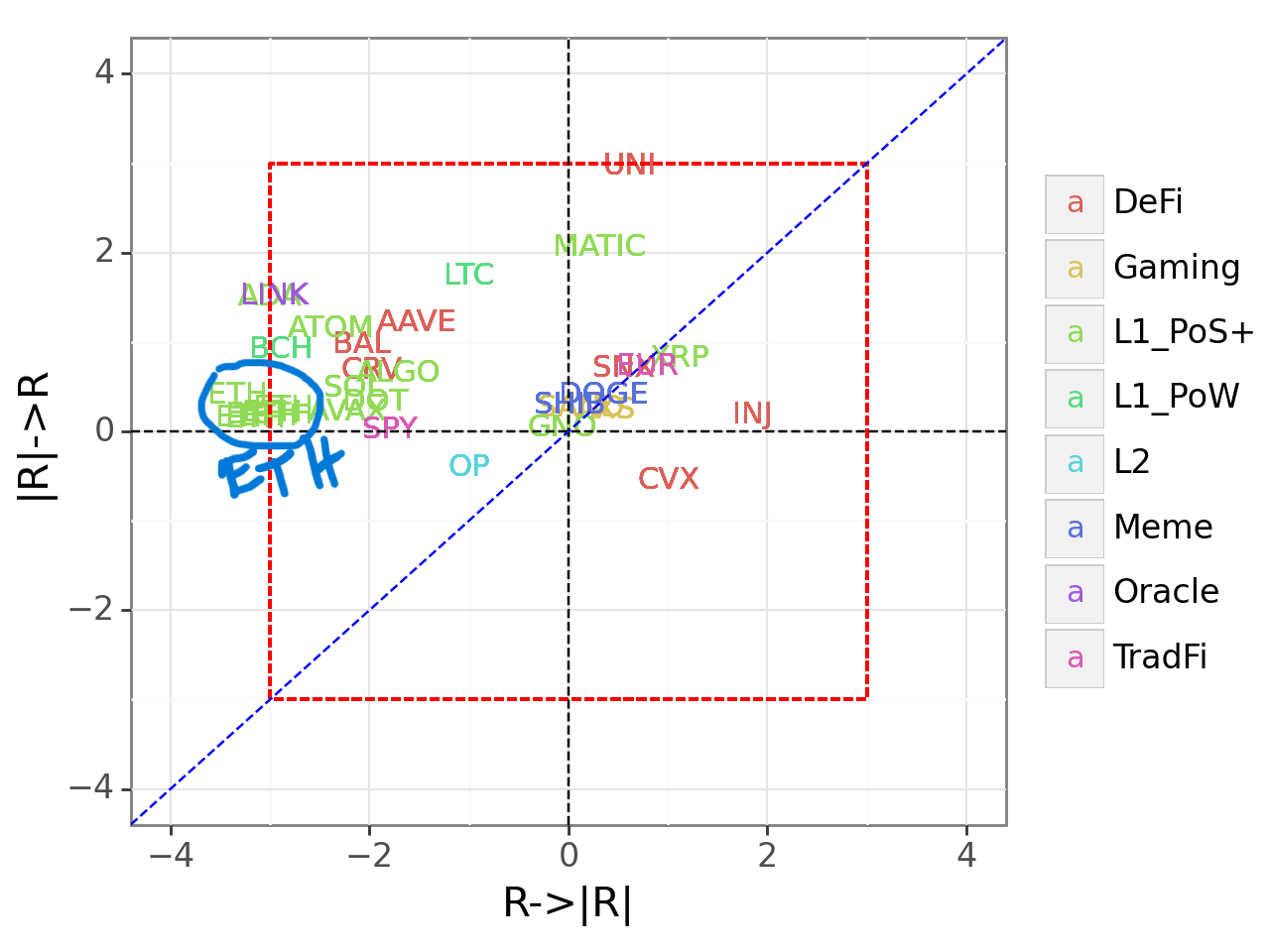}
\caption{The horizontal axis shows the average cross-correlation [L(k)] between the past returns and future absolute returns (labeled R to the absolute of R). The vertical axis shows the average cross-correlation [L(k)] of past absolute returns and future returns. An asset shows the leverage effect if it is on the negative x-axis in this figure. A typical example is the SPY, which is on top of the horizontal axis on the negative quadrant. We also include $\pm 3$ (99\%) error bars as a red dashed box. Values inside the box are insignificant for both $k>0$ or $k<0$.}
\label{fig:lev_stat2}
\end{figure}

To condense this information, Figure \ref{fig:lev_stat2} plots the average $L(k)$ for $k<0$ against the average $L(k)$ for $k>0$. The average for $k<0$ (labeled "$R \to |R|$") measures the average strength of the relation between current and future absolute returns (a proxy for volatility). The leverage effect expects this relation to be negative: a large price drop (negative return) is followed by periods of increased volatility (absolute returns). It also expects the average $L(k)$ for $k>0$ (labeled "$|R| \to R$") to be close to zero, indicating that current absolute returns do not forecast future returns.

In Figure \ref{fig:lev_stat2}, SPY exemplifies the leverage effect on the negative side of the x-axis. ETH forms a cluster from different exchanges close to the x-axis at negative values often larger than $-3$ (red dashed box), though the distance to this x-axis varies among ETH data sources.

The remaining tokens appear to form two main groups with some outliers. One group (BCH, SOL, ARGO, etc.) is close to ETH and SPY, exhibiting the leverage effect. The other group, including EUR, GNO, Meme, and Gaming coins, is closer to the origin and along the diagonal, not showing the leverage effect. Outliers such as UNI, INJ, CVX, MATIC, OP, and LTC are challenging to classify due to large error bars. Notably, UNI sits on the red dashed line close to the y-axis, suggesting a possible reverse leverage effect.

\subsection{Time Reversal Asymmetry}
Time reversal asymmetry (TRA) in financial time series, introduced by \cite{Z2009}, is a concept linked to entropy and time directionality. It suggests that financial time series exhibit different statistical features when viewed in reverse chronological order. This phenomenon can be illustrated by the analogy of observing broken glass: while one can easily deduce that glass fragments originated from an intact cup, it's nearly impossible to envision the original cup when presented only with its fragments.

In financial terms, this analogy translates to the day's volatility, measured as the absolute value of the open-to-close return (the "full glass"), being a better predictor of future daily volatility, computed as the standard deviation of intraday returns (the "broken pieces"), than vice versa.

To quantify TRA, we employ a correlation measure introduced in \cite{Rose2020}:
\begin{equation} \label{eq:TRA}
C(k) \equiv
\frac{ \mathbb{E}\left[ ( \lvert R_t \rvert - \mu_{\lvert R \rvert} )
( s^{(D)}_{t+k}  - \mu_{s^{(D)}} )\right] }{ \sigma_{\lvert R \rvert} \sigma_{s^{(D)} }}
\end{equation}

\noindent where $R$ is the daily return (absolute value of open to close return) and $s^{(D)}$ is the empirical intraday volatility (standard deviation of hourly returns within a day).

TRA is observed when $C(k) \geq C(-k)$, indicating that the predictability of intraday volatility using absolute return surpasses the reverse. To visualize this asymmetry, we compute the cumulative difference:
\begin{equation} \label{eq:TRA cumdiff}
\Delta(N) = \sum^N_{k=1} \left[ C(k) - C(-k) \right]
\end{equation}

\noindent An increasing function of $\Delta(N)$ confirms the presence of TRA \cite{Rose2020}.

\begin{figure}[ht]
\centering
\includegraphics[width=1\linewidth]{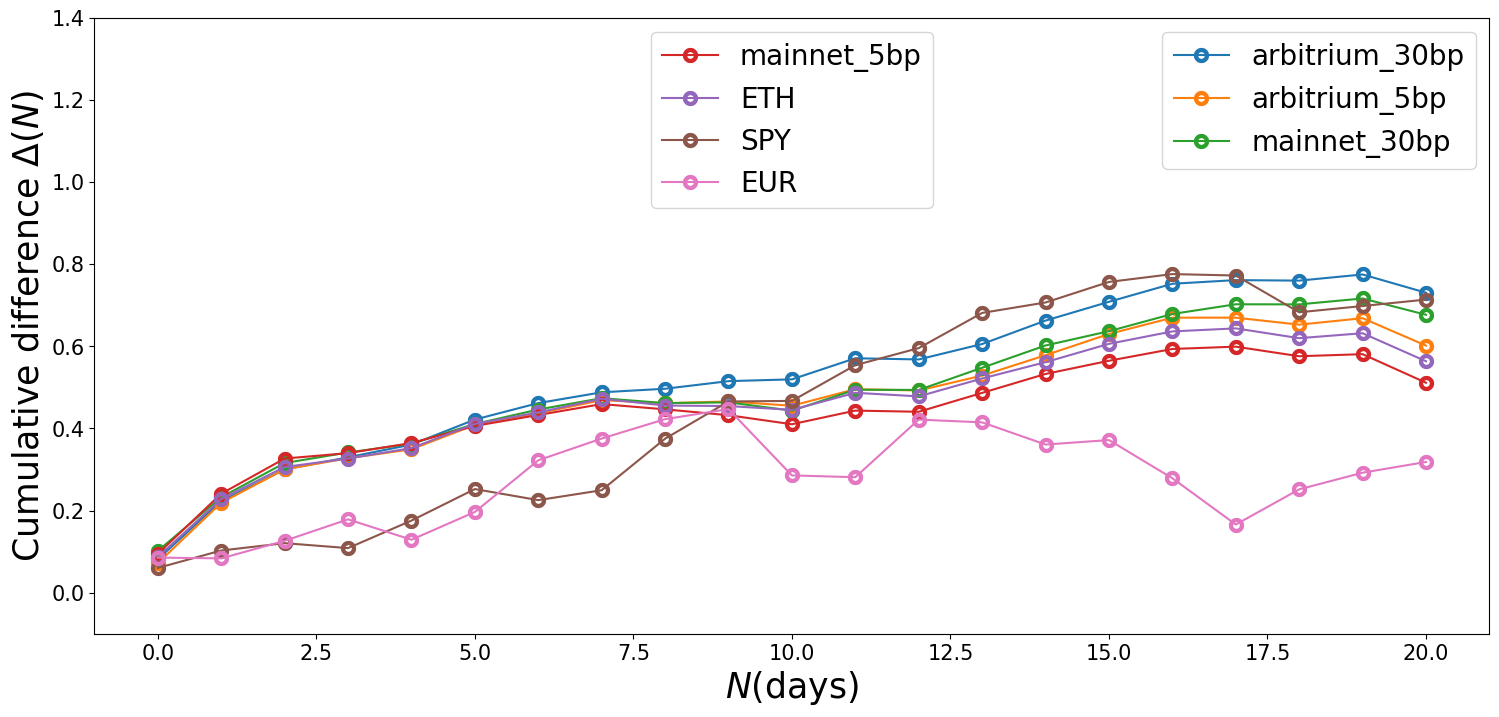}
\caption{Cumulative difference function $\Delta(N)$ (Equation~\eqref{eq:TRA cumdiff}) for ETH in different exchanges as well as SPY and EUR.}
\label{fig:pool_CEX_TRA}
\end{figure}

Figure \ref{fig:pool_CEX_TRA} illustrates $\Delta(N)$ for selected assets over the same historical period. The consistently increasing lines for all assets, including both crypto and traditional financial instruments (SPY and EUR), validate the presence of TRA, aligning with previous findings in traditional finance \cite{Z2009, Rose2020}.

ETH exhibits TRA comparable to SPY and EUR, regardless of the exchange type (DEX or CEX). However, similar to the leverage effect, we find that TRA is not universally present across all studied tokens.

\begin{figure}[ht]
\centering
\includegraphics[width=1\linewidth]{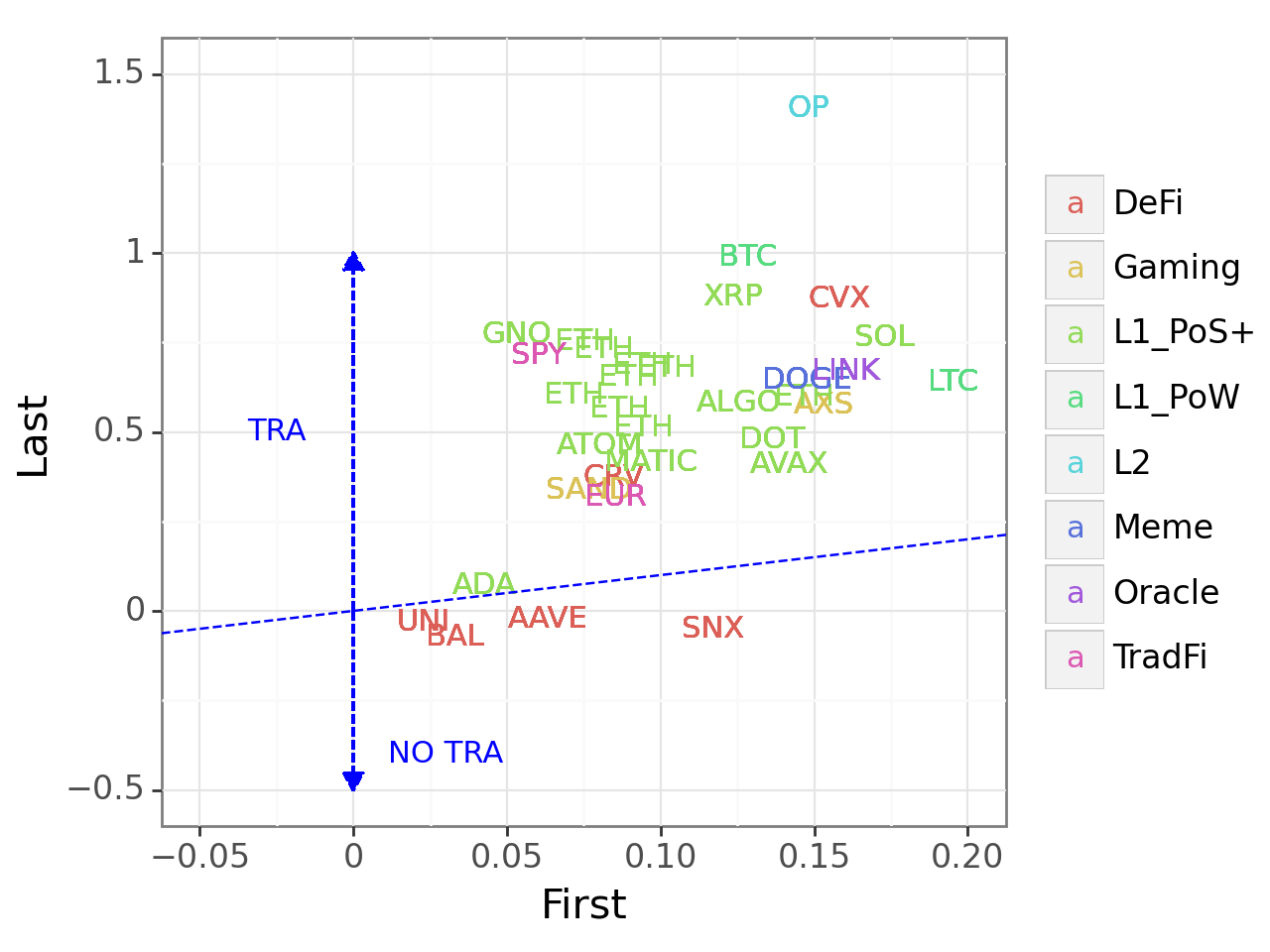}
\caption{Scatter plot of $\Delta(N=1)$ vs $\Delta(N=20)$ for all tokens, color-coded by sector. The blue dashed diagonal line separates regions of TRA (above) and no TRA (below).}
\label{fig:TRA}
\end{figure}

To provide a comprehensive overview of TRA across our dataset, Figure \ref{fig:TRA} plots $\Delta(N=1)$ against $\Delta(N=20)$ for all tokens, color-coded by sector. This representation compresses the information by focusing on the first and last terms of the cumulative sum. The presence of TRA is indicated by data points above the diagonal line, where the last element of the sum exceeds the first.

Our analysis reveals that most tokens exhibit TRA consistent with traditional assets. However, there are notable exceptions. Interestingly, two tokens associated with the largest DEX, UNI and BAL, do not demonstrate TRA.

\subsection{Correlation Matrix and Market Factors}
To analyze the cross-section of the token market, we expanded our focus group from 25 to 145 tokens. Data was sourced from Coinbase, a centralized exchange, starting from 2022-03-16. We down-sampled hourly returns to daily returns to facilitate comparison with prevalent stock market studies.

\begin{figure}[ht]
\centering
\includegraphics[width=1\linewidth]{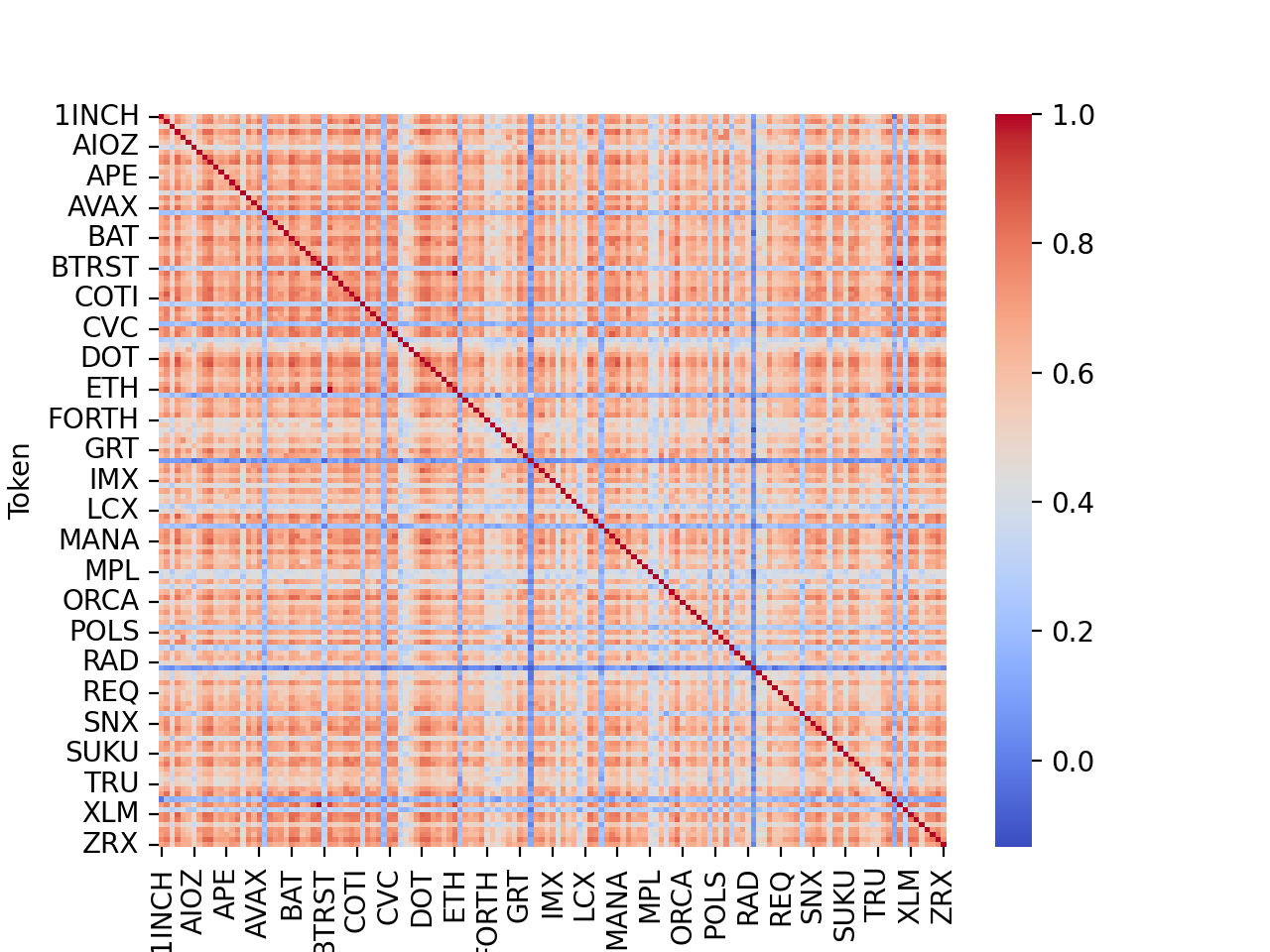}
\caption{Correlation matrix for 145 tokens downloaded from Coinbase. The correlation was calculated using daily returns starting in 2022-03.}
\label{fig:cor_mat}
\end{figure}

Figure \ref{fig:cor_mat} presents a heat map of the correlation matrix of daily returns for these 145 tokens. This set represents the largest available list on Coinbase, traded hourly since 2022 without interruption. Our selection criteria prioritized the most comprehensive list of tokens with consistent data over the past two years without applying additional filters. The visualization demonstrates that a majority of these tokens exhibit high inter-correlations.

For comparison with the stock market, we gathered two years of daily closing prices for about 1,900 US stocks from Yahoo Finance, corresponding to our token history timeline. We filtered this stock data by including only stocks priced above \$10 on the data collection day and excluding those with daily returns correlated more than 97\% with any other stock. This approach eliminated duplicate company listings and other data anomalies.

\begin{figure}[ht]
\centering
\includegraphics[width=1\linewidth]{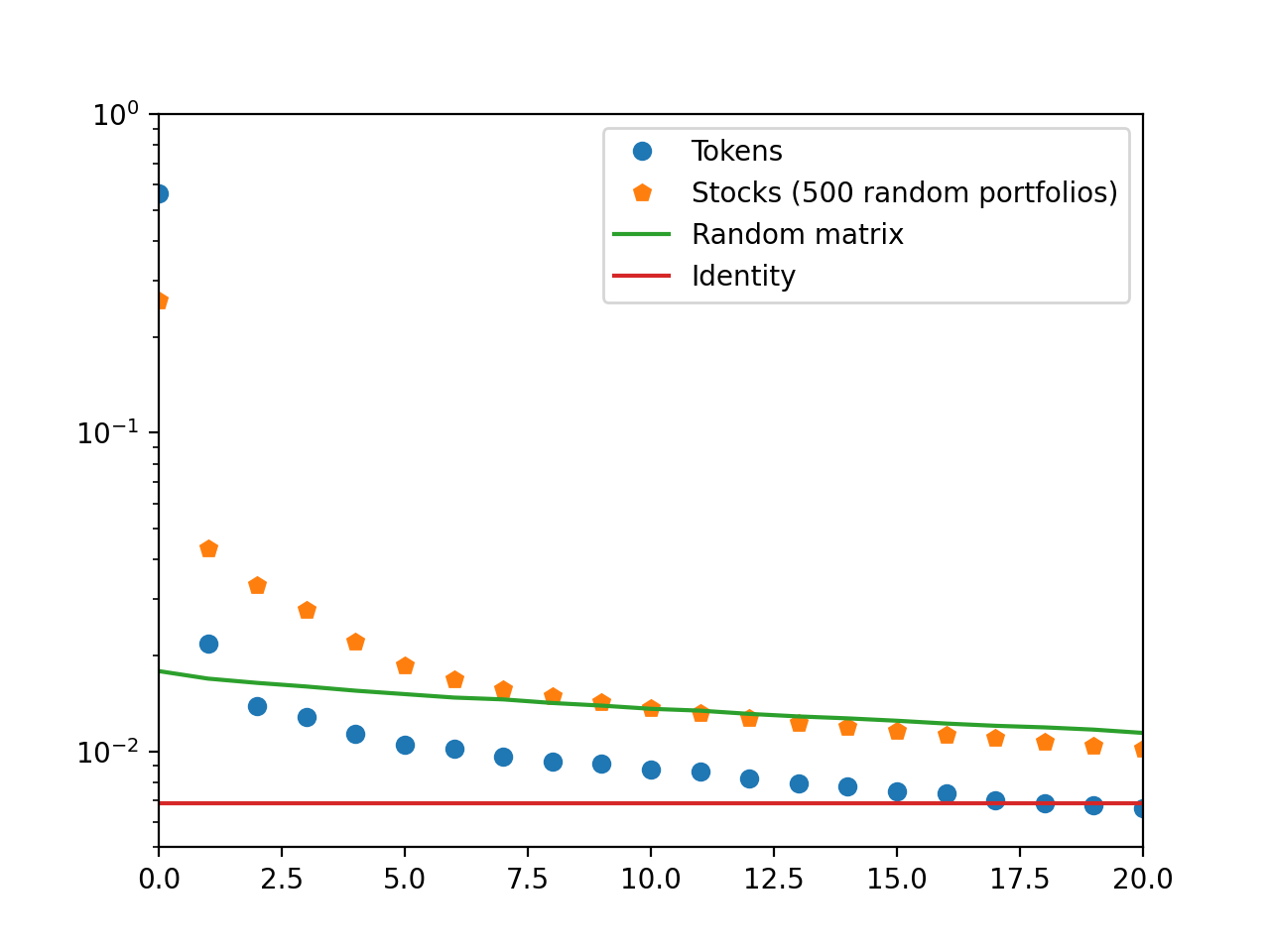}
\caption{First top 20 eigenvalues for the token universe (blue) and Stocks (orange). The solid green line is the eigenvalue spectrum for a random matrix. Eigenvalues above the green line could be significant, particularly the first eigenvalue.}
\label{fig:eigen}
\end{figure}

Our analysis randomly sampled 145 stocks from this 1,900-stock universe 500 times with replacements. For each sample, we calculated eigenvalues and then averaged these across all samples. The eigenvalues were sorted in descending order and normalized by dividing each by the sum of all eigenvalues. Figure \ref{fig:eigen} presents the top 20 eigenvalues from this analysis.

Figure \ref{fig:eigen} reveals striking differences between the token and stock markets. In the token universe, the top eigenvalue accounts for 56\% of the total variance, whereas in the stock samples, it explains only about 26\% (reported conservatively as the average minus three standard errors over the 500 randomly selected stock portfolios). Importantly, the first eigenvector is positive for all tokens in our sample, consistent with its interpretation as the "market" factor, as discussed in \cite{PLEROU2000374,Cont2001,Bouchaud_Potters_2003}.

Moreover, Figure \ref{fig:eigen} shows that only two eigenvalues appear significant for the token universe, exceeding the threshold expected for randomly uncorrelated assets (green line). In contrast, stocks exhibit at least six significant eigenvalues.

These findings suggest that the token universe is driven by a smaller set of eigenvalues/factors than the stock market while still adhering to the stylized fact of factor-driven markets observed in stocks \cite{PLEROU2000374,Cont2001,Bouchaud_Potters_2003}. This conclusion has important implications for modeling and risk control of token portfolios. Specifically, it suggests that a few factor models—likely fewer than those required for stocks—may be sufficient to effectively model the cross-section of tokens.

\subsubsection{Correlation Across Time}
While illustrating the temporal evolution of the correlation matrix presents challenges due to limited reliable data, we can gain qualitative insights by analyzing a subset of our dataset. We focus on 25 tokens with data beginning in September 2020, extending our analysis two years earlier than the data used in Figure \ref{fig:eigen}. This period encompasses significant market phases, including a bull market that saw Bitcoin's price increase nearly five-fold, followed by substantial volatility in early 2021, and eventually establishing a bull market toward the end of 2021.

To analyze correlation dynamics across these market regimes, we examine the time series of the first eigenvalue of the correlation matrix for our 25-token subset. Our methodology involves calculating a 60-day trailing correlation matrix daily, computing its eigenvalue spectrum, and recording the fraction of variance explained by the first eigenvalue at the end of each 60-day period.

\begin{figure}[ht]
\centering
\includegraphics[width=1\linewidth]{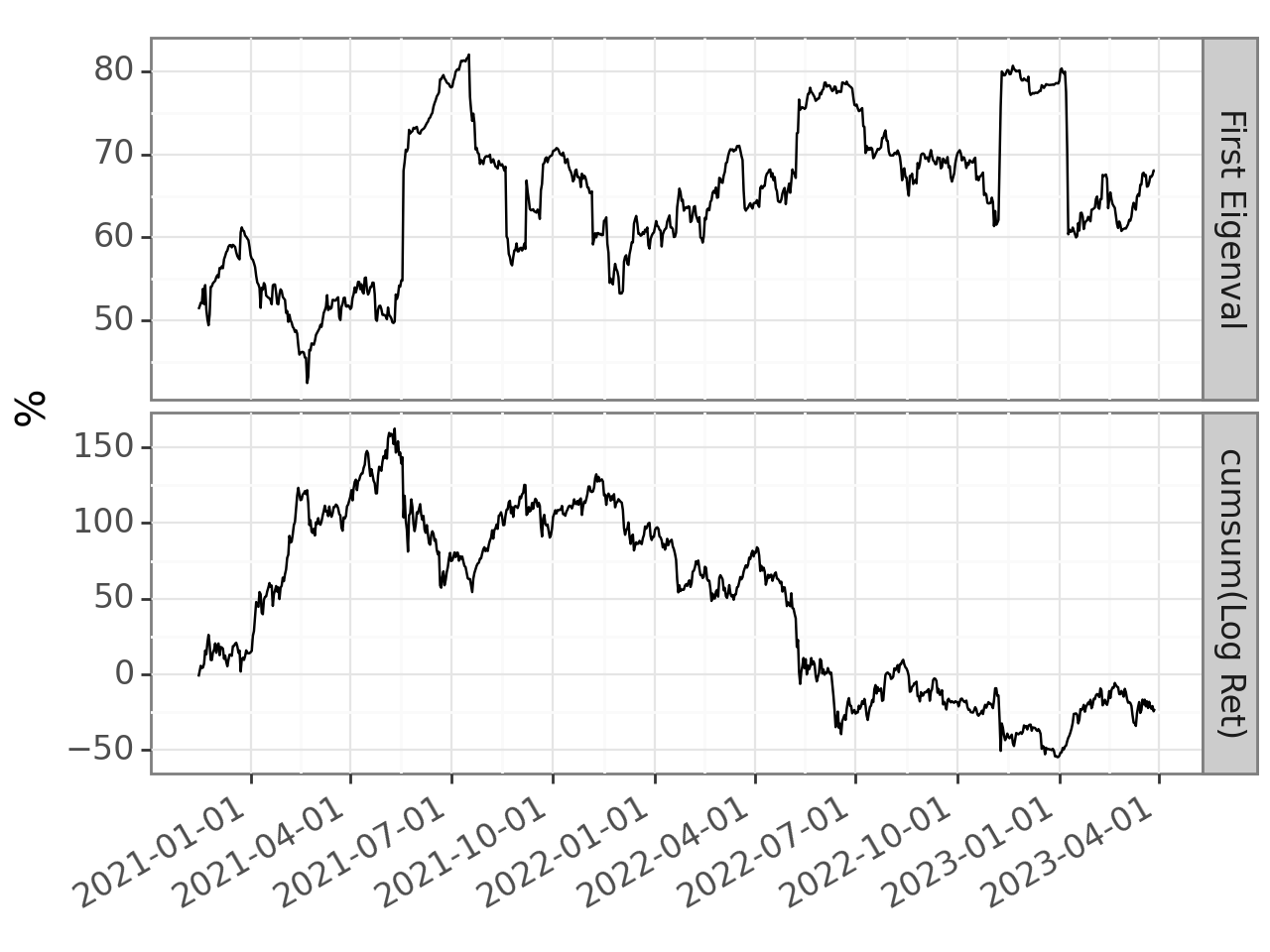}
\caption{Time-series of the variance explained by the first eigenvalue for a rolling 60-day correlation matrix of 25 tokens (top) with the cumulative sum of the log-returns for the same portfolio of 25 tokens (lower). The graph illustrates that bull markets are more diversified (less variance explained by the first eigenvalue) and bear markets are less diversified (more variance explained by the first eigenvalue).}
\label{fig:eigenTS}
\end{figure}

Figure \ref{fig:eigenTS} presents two key elements: the time series of the fraction of variance explained by the first eigenvalue (upper panel) and the cumulative sum of log-returns averaged across our 25-token sample (lower panel). During the bear period after 2022, the average explained variance approaches 66\%, notably higher than the 56\% observed in our previous section. This difference stems from our use of a substantially smaller, less diversified correlation matrix in this analysis.

Despite the limited data span, our analysis reveals distinct patterns in market behavior. The first eigenvalue tends to be smaller (approximately 50\%) during bull markets and higher (around 65\%) during bear markets. This transition became particularly evident during the significant market downturn in 2021. Subsequently, the fraction of explained variance has fluctuated around 65\%, occasionally reaching 80\% during rapid market declines. These findings demonstrate a negative relationship between token market performance (particularly Bitcoin) and the fraction of explained variance. During substantial market upswings, tokens show more independent movement, while market crashes tend to trigger synchronized downward movements across tokens.

This breakdown in diversification benefits during market crashes mirrors well-documented patterns in traditional stock markets \cite{preis2012quantifying}. Our analysis suggests that Web3 tokens exhibit similar behavior, where diversification benefits diminish precisely when they would be most valuable – during market downturns.

\subsection{Token Statistics and Characteristics}
Table \ref{tab:asset_table} provides comprehensive statistics for the tokens studied in this article. The "Token" column lists the tokens and exchange information, including Binance and Coinbase as large centralized exchanges, and "arb" and "m" denoting pools within Uniswap, the largest decentralized spot exchange. Here, "arb" stands for Arbitrum, an L2 chain solution on top of the Ethereum L1 blockchain, while "m" represents Mainnet, the Ethereum L1 chain.

The "Sector" column classifies the different tokens according to their functionalities, with a visual representation in Figure \ref{fig:tok}. The "History" column presents each token's historical price data range used in this work, while the "Zeros(\%)" column shows the fraction of zero one-hour returns for each token given the available history.

We quantify various financial phenomena and statistical properties in the subsequent columns. The "Avg ACF" column quantifies the lack of autocorrelation of returns, except at specific lags. It presents the average autocorrelation of returns at key points in the autocorrelation spectrum, with "1+24" showing the average for the first and 24th hours and "Else" representing the average for all lags until hour 96, excluding hours 1 and 24. Figure \ref{fig:Pool_CEX_trad_acf_R} illustrates a typical return autocorrelation plot, while Figure \ref{fig:acf_sum} visually represents the values in the table.

Volatility clustering is quantified in the "ACF abs(R)" column, which shows the intercept and slope of a linear regression between the log of the autocorrelation lag in hours and the log of the ACF of the absolute return (refer to Figure \ref{fig:Pool_CEX_trad_acf_absR}). The "Avg Lev" column quantifies the Leverage effect, presenting the average of the negative ("-") and positive ("+") branches of the cross-correlation between return and absolute return. Figure \ref{fig:pool_CEX_lev} demonstrates a typical Leverage effect, with Figure \ref{fig:lev_stat2} offering a visual representation of the values in the table except for SPY and EUR. SPY and EUR values in the table refer to the entire history, starting in 1998 for SPY and 2002 for EUR. However, Figure \ref{fig:lev_stat2} shows "Avg Leg" for SPY and EUR with historical data starting in 2021 in sync with the pool historical data. The table shows SPY with a significant leverage effect (-8.4).

Time reversal asymmetry is captured in the "TRA" column, which presents the initial cumulative difference ("Ini") and the final cumulative difference ("Fin"). Figure \ref{fig:pool_CEX_TRA} illustrates a typical TRA, while Figure \ref{fig:TRA} visually represents these values. The "CDF Tail" column quantifies the fatness of power tails, showing exponents of a power law fit to the cumulative density of returns for both positive ("R") and negative ("L") tails of normalized returns after two standard deviations. Detailed explanations and illustrative figures are in Appendix \ref{app:tail}.

Lastly, the "JB" column quantifies aggregated normality by presenting the slope of a linear fit to the log number of days and the log of the JB statistic, as illustrated in Figure \ref{fig:pool_CEX_JB_stats}.

\begin{table}[p]  
\centering
\scriptsize  
\setlength{\tabcolsep}{4pt}  
\begin{threeparttable}
\resizebox{\textwidth}{!}{
\begin{tabular}{l l l r *{4}{c} *{2}{r} *{2}{c} *{2}{c} c}
\toprule
\multirow{2}{*}{Token} & \multirow{2}{*}{Sector} & \multirow{2}{*}{History} & \multirow{2}{*}{Zeros(\%)} & 
\multicolumn{2}{c}{Avg ACF} & \multicolumn{2}{c}{ACF Abs(R)} & \multicolumn{2}{c}{Avg Lev} & \multicolumn{2}{c}{TRA} & \multicolumn{2}{c}{CDF Tail} & \multirow{2}{*}{JB} \\
\cmidrule(lr){5-6} \cmidrule(lr){7-8} \cmidrule(lr){9-10} \cmidrule(lr){11-12} \cmidrule(lr){13-14}
& & & & 1+24 & Else & Slope & Int. & + & - & Ini & Fin & R & L & \\
\midrule
SPY & TradFi & 1998/7/1--2023/12/5 & 2 & -0.86 & 3.9 & NaN & NaN & 0.52 & -8.4 & 0.048 & 0.34 & 2.5 & 2.5 & -1.5 \\
EUR & TradFi & 2002/10/21--2023/12/5 & 2 & -0.82 & 3.8 & NaN & NaN & 0.1 & 0.3 & 0.091 & 0.41 & 2.7 & 2.7 & -1.8 \\
Binance BTC & PoW+L1 & 2021/9/23--2023/10/18 & $<$1 & -1.3 & 1.6 & -0.26 & -1.3 & 0.4 & -1.6 & 0.13 & 0.99 & 2.4 & 2.5 & -2.3 \\
Binance XRP & PoS+L1 & 2020/12/31--2024/1/31 & 1 & -4 & 1.7 & -0.22 & -1.2 & 0.82 & 1.1 & 0.12 & 0.88 & 2.3 & 2.4 & -3 \\
Binance ETH & PoS+L1 & 2017/8/17--2023/10/18 & 1 & -9.1 & 2.5 & -0.22 & -1.2 & 1.5 & -6.3 & 0.21 & 0.23 & 2.6 & 2.5 & -1.8 \\
$\text{arb}_{30bp}$ DAI & L1 & 2021/9/23--2023/10/18 & 12 & 1.5 & 1.5 & -0.21 & -1.2 & 0.2 & -3 & 0.075 & 0.75 & 2.5 & 2.5 & -1.5 \\
$\text{arb}_{30bp}$ USDC & L1 & 2021/9/23--2023/10/18 & 5 & 0.12 & 1.5 & -0.19 & -1.1 & 0.18 & -2.9 & 0.081 & 0.73 & 2.6 & 2.5 & -1.5 \\
$\text{arb}_{5bp}$ USDC & L1 & 2021/9/23--2023/10/18 & $<$1 & -1.1 & 1.6 & -0.2 & -1.1 & 0.27 & -2.9 & 0.072 & 0.6 & 2.5 & 2.5 & -1.5 \\
$\text{arb}_{5bp}$ USDT & L1 & 2021/9/23--2023/10/18 & $<$1 & -1.7 & 1.5 & -0.22 & -1.1 & 0.41 & -3.3 & 0.15 & 0.6 & 2.5 & 2.4 & -1.7 \\
Coinbase AAVE & Saving+Defi & 2020/12/15--2023/11/6 & $<$1 & -2.7 & 1.7 & -0.15 & -1.1 & 1.2 & -1.5 & 0.063 & -0.024 & 2.8 & 2.9 & -1.6 \\
Coinbase ADA & PoS+L1 & 2021/3/18--2023/11/6 & 2 & -9.1 & 1.6 & -0.19 & -1.3 & 1.5 & -3 & 0.043 & 0.07 & 2.6 & 2.6 & -2 \\
Coinbase ALGO & PoS+L1 & 2019/8/15--2023/11/6 & 3 & -8.5 & 2 & -0.19 & -1.3 & 0.64 & -1.7 & 0.13 & 0.58 & 2.9 & 3 & -1.9 \\
Coinbase ATOM & PoS+L1 & 2020/1/14--2023/11/6 & $<$1 & -7.4 & 1.9 & -0.18 & -0.98 & 1.2 & -2.4 & 0.08 & 0.46 & 2.7 & 2.9 & -2.3 \\
Coinbase AVAX & PoS+L1 & 2021/9/30--2023/11/6 & 4 & -5.8 & 1.5 & -0.2 & -1 & 0.22 & -2.2 & 0.14 & 0.41 & 2.8 & 2.8 & -2.1 \\
Coinbase AXS & Gaming & 2021/8/12--2023/11/6 & 6 & -5.1 & 1.4 & -0.24 & -1.2 & 0.24 & 0.38 & 0.15 & 0.57 & 2.2 & 2.7 & -2 \\
Coinbase BAL & Exchange+Defi & 2020/10/6--2023/11/6 & 7 & -3.7 & 1.7 & -0.16 & -1.4 & 0.97 & -2.1 & 0.033 & -0.075 & 3.2 & 2.9 & -1.8 \\
Coinbase BCH & PoW+L1 & 2017/12/20--2023/11/6 & $<$1 & -7.9 & 2.6 & -0.27 & -0.87 & 0.91 & -2.9 & 0.21 & 0.74 & 2.4 & 2.5 & -2.7 \\
Coinbase BTC & PoW+L1 & 2016/1/1--2023/11/6 & $<$1 & -8.3 & 2.8 & -0.2 & -1.1 & 1.3 & -6.2 & 0.23 & 0.48 & 2.5 & 2.4 & -2.3 \\
Coinbase CRV & Exchange+Defi & 2021/3/25--2023/11/6 & 1 & -2.4 & 1.5 & -0.19 & -0.92 & 0.68 & -2 & 0.085 & 0.37 & 2.8 & 2.8 & -2.1 \\
Coinbase CVX & Stacking+Defi & 2022/9/8--2023/11/6 & 18 & -2.5 & 0.99 & -0.33 & -1.5 & -0.55 & 1 & 0.16 & 0.87 & 2.5 & 2.5 & -1.9 \\
Coinbase DOGE & PoS+Memecoin & 2021/6/3--2023/11/6 & 2 & -5.3 & 1.5 & -0.33 & -1.4 & 0.4 & 0.36 & 0.15 & 0.64 & 2.3 & 2.6 & -2.1 \\
Coinbase DOT & PoS+L1 & 2021/6/16--2023/11/6 & 1 & -3.5 & 1.4 & -0.23 & -1 & 0.33 & -1.9 & 0.14 & 0.48 & 2.7 & 2.8 & -1.9 \\
Coinbase ETH & PoS+L1 & 2016/5/18--2023/11/6 & 1 & -7.6 & 2.9 & -0.24 & -1.1 & 2.7 & -5.6 & 0.21 & 0.56 & 2.5 & 2.5 & -1.8 \\
Coinbase GNO & PoS+L1 & 2022/7/21--2023/11/6 & 24 & -11 & 0.99 & -0.35 & -0.96 & 0.041 & -0.063 & 0.053 & 0.77 & 3.3 & 3.3 & -1.7 \\
Coinbase INJ & Exchange+Defi & 2022/9/20--2023/11/6 & 3 & -4.1 & 1.2 & -0.19 & -1.2 & 0.19 & 1.9 & 0.043 & -0.55 & 2.7 & 2.7 & -1.6 \\
Coinbase LINK & Oracle & 2019/6/27--2023/11/6 & $<$1 & -9.1 & 2.1 & -0.23 & -1.5 & 1.5 & -3 & 0.16 & 0.67 & 2.9 & 2.7 & -2.2 \\
Coinbase LTC & PoW+L1 & 2016/8/17--2023/11/6 & 4 & -11 & 2.7 & -0.25 & -0.8 & 1.7 & -1 & 0.2 & 0.64 & 2.5 & 2.6 & -1.6 \\
Coinbase MATIC & L2 & 2021/3/11--2023/11/6 & $<$1 & -6.7 & 1.7 & -0.15 & -1.2 & 2.1 & 0.31 & 0.097 & 0.42 & 2.5 & 2.8 & -2.2 \\
Coinbase OP & L2 & 2022/6/1--2023/11/6 & 4 & -0.15 & 1.3 & -0.19 & -0.82 & -0.41 & -1 & 0.15 & 1.4 & 2.5 & 2.6 & -1.8 \\
Coinbase SAND & Gaming & 2022/5/26--2023/11/6 & 2 & -0.91 & 1.2 & -0.21 & -1.1 & 0.26 & 0.13 & 0.077 & 0.34 & 2.7 & 2.6 & -1.8 \\
Coinbase SHIB & PoS+Memecoin & 2022/1/1--2023/11/6 & 8 & -7.4 & 1.4 & -0.34 & -1.2 & 0.29 & 0.015 & 0.2 & 0.71 & 2.3 & 2.5 & -2.6 \\
Coinbase SNX & Exchange+Defi & 2020/12/15--2023/11/6 & 2 & -5.9 & 1.5 & -0.22 & -1.2 & 0.7 & 0.55 & 0.12 & -0.05 & 2.8 & 3 & -2.3 \\
Coinbase SOL & PoS+L1 & 2021/6/17--2023/11/6 & 2 & -4.1 & 1.8 & -0.33 & -1.2 & 0.49 & -2.2 & 0.17 & 0.76 & 2.6 & 2.9 & -1.8 \\
Coinbase UNI & Exchange+Defi & 2020/9/17--2023/11/6 & 1 & -5.1 & 1.6 & -0.19 & -1.2 & 3 & 0.62 & 0.023 & -0.032 & 2.7 & 2.8 & -1.7 \\
$\text{m}_{30bp}$ DAI & L1 & 2021/9/23--2023/10/18 & 19 & -0.21 & 1.5 & -0.19 & -1.2 & 0.18 & -3.1 & 0.094 & 0.69 & 2.5 & 2.5 & -1.6 \\
$\text{m}_{30bp}$ USDC & L1 & 2021/9/23--2023/10/18 & 8 & -0.46 & 1.5 & -0.18 & -1.2 & 0.15 & -3.2 & 0.1 & 0.68 & 2.4 & 2.5 & -1.5 \\
$\text{m}_{30bp}$ USDT & L1 & 2021/9/23--2023/10/18 & 4 & -0.64 & 1.6 & -0.18 & -1.2 & 0.14 & -3.1 & 0.089 & 0.65 & 2.5 & 2.5 & -1.5 \\
$\text{m}_{5bp}$ USDC & L1 & 2021/9/23--2023/10/18 & 6 & -1.8 & 1.6 & -0.18 & -1.2 & 0.17 & -3 & 0.095 & 0.51 & 2.4 & 2.5 & -1.5 \\
\bottomrule
\end{tabular}
}
\end{threeparttable}
\caption{Assets and statistics. We call "arb" the Arbitrum pool and "m" the Mainnet pool.}
\label{tab:asset_table}
\end{table}

\section{Discussion and Conclusion}

\subsection{Overview of Stylized Facts in the Cryptocurrency Market}
Our comprehensive analysis reveals that many stylized facts well-documented in equity markets are also present in the cryptocurrency token universe. Fat tails, aggregate normality, and volatility clustering are ubiquitous across all analyzed tokens. However, the leverage effect and time reversal asymmetry are less prevalent. The correlation structure of the token market adheres to the few-factor rule observed in equity markets.

Interestingly, the trading venue (DEX vs. CEX) does not significantly impact these stylized facts despite substantial differences in liquidity and trading mechanisms. For instance, ETH exhibits consistent stylized facts across different trading platforms.

Our analysis across different token sectors reveals that the consensus mechanism (PoW vs. PoS) or the classification as L1 or L2 does not qualitatively alter the presence of stylized facts. However, we observed that gaming and meme tokens tend not to exhibit the leverage effect, although this finding is subject to statistical uncertainty.

\subsection{Quantitative Differences Among Tokens}
Despite qualitative similarities in stylized facts, quantitative differences exist among tokens. Table \ref{tab:asset_table} presents a series of statistics that provide a more nuanced picture of these differences.

Many tokens exhibit substantially different data histories, stemming from data access limitations and the varying lifespans of projects. For instance, Bitcoin (BTC) on Coinbase has the longest history, being the first cryptocurrency created. While the length of historical data doesn't alter the presence of stylized facts qualitatively, it does impact quantitative measurements such as autocorrelation values and tail exponents. It's important to note that these quantitative differences don't fundamentally change our conclusions; for example, variations in tail exponents are not significant enough to transform a power law distribution into an exponential one.

The fraction of zero one-hour returns (column Zero(\%) in Table \ref{tab:asset_table}) serves as a reliable liquidity measure. BTC and ETH demonstrate high liquidity, with less than 1\% zero returns, reflecting their dominance in the cryptocurrency market (over 70\% market share). They appear more liquid than traditional assets like SPY and EUR, though differences in price rounding precision complicate this comparison.

Ethereum traded on Uniswap's 30 basis point (bp) pools (both Mainnet and Arbitrum) shows significantly lower liquidity, with over 50\% zero ten-minute returns (19\% hourly returns) compared to just 4\% for the 5bp pool (<1\% hourly returns). This stark difference indicates infrequent price movements and low trading activity in the 30bp pools.\footnote{The Mainnet USDT pool (trading ETH against USDT on the L1 ETH blockchain) emerges as the most liquid 30bp pool among stablecoin pairs}. The liquidity disparity between 30bp and 5bp pools is largely attributable to trading costs. With 30bp pools charging six times more than 5bp pools for the same trade, traders naturally gravitate towards the more cost-effective option.

Despite their obvious disadvantages, the continued existence of 30bp pools offers insights into DeFi's decentralized nature. These pools, created before their 5bp counterparts, persist due to the absence of a centralized authority capable of decommissioning them. Their gradual phasing out is expected as traders and liquidity providers naturally migrate to more efficient pools. This persistence of less efficient structures in DeFi environments highlights the ecosystem's evolutionary nature, where market forces, rather than centralized decisions, drive changes in liquidity distribution and trading practices. It also underscores the importance of considering historical context and decentralized governance when analyzing DeFi market structures.

\subsection{Time Stability of Stylized Facts}
The statistics studied in this article can exhibit significant temporal variation. We examine this variation through BTC and ETH, focusing on two distinct temporal frameworks: market regimes and the May 2020 halving event.

\subsubsection{Market Regime Analysis} \label{sec:reg}
To deepen our understanding of token market dynamics, we compare the stylized facts of BTC and ETH in two market regimes after 2020. This period is particularly interesting as it encompasses two contrasting market phases: a significant bull market from March 2020 to April 2021 and a pronounced bear market from December 2021 to January 2023.

\begin{figure}[ht]
\centering
\includegraphics[width=1\linewidth]{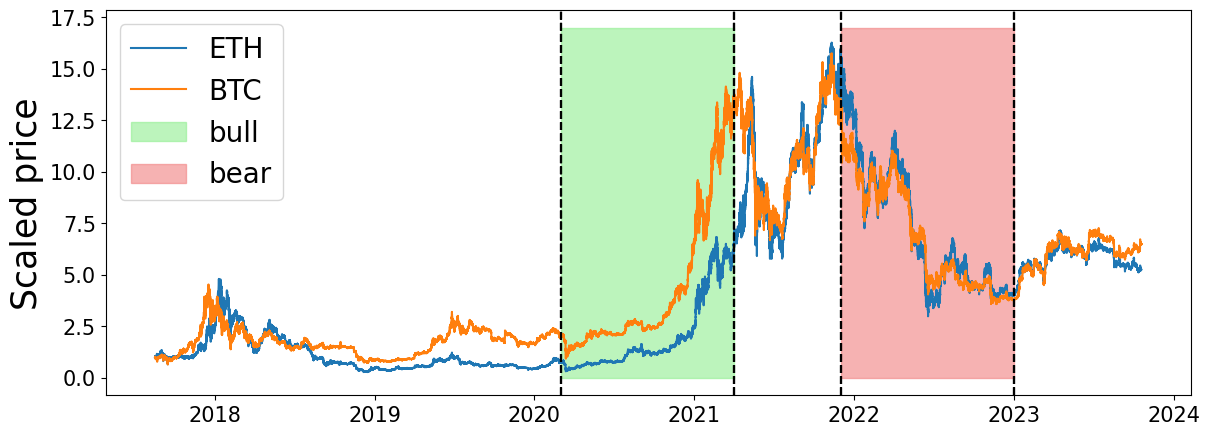}
\caption{Scaled price of ETH and BTC starting from June 2017. The graph illustrates a bull market and a bear market after 2020.}
\label{fig:regime}
\end{figure}

Figure \ref{fig:regime} illustrates these distinct market phases through the scaled price movements of both tokens. The contrast between the bull and bear markets provides an excellent natural experiment for examining how market conditions affect the statistical properties of these assets.

Our empirical analysis reveals several key findings. Table \ref{table:regimes} shows that certain fundamental properties remain stable across market regimes. The unconditional return distribution, measured by the tail of the cumulative distribution function (CDF Tail), and volatility clustering (ACF Abs(R)) demonstrate remarkable consistency between regimes. Both ETH and BTC exhibit similar patterns in these metrics. While these findings broadly align with Table \ref{tab:asset_table}, we observe an important distinction in the absolute return autocorrelation. The full dataset shows a faster decay with a slope of approximately $-0.22$ (Table \ref{tab:asset_table}), compared to $-0.13$ in our regime analysis (Table \ref{table:regimes}). Section \ref{sec:half} attributes this difference to structural changes around the May 2020 halving event.

\begin{table}[ht]
\centering
\small
\resizebox{\textwidth}{!}{
\begin{tabular}{l c c r *{4}{c} *{2}{r} *{2}{c} *{2}{c} c}
\toprule
\multirow{2}{*}{Token} & \multirow{2}{*}{Regime} & \multirow{2}{*}{History} & \multirow{2}{*}{Zeros(\%)} & 
\multicolumn{2}{c}{Avg ACF} & \multicolumn{2}{c}{ACF Abs(R)} & \multicolumn{2}{c}{Avg Lev} & \multicolumn{2}{c}{TRA} & \multicolumn{2}{c}{CDF Tail} & \multirow{2}{*}{JB} \\
\cmidrule(lr){5-6} \cmidrule(lr){7-8} \cmidrule(lr){9-10} \cmidrule(lr){11-12} \cmidrule(lr){13-14}
& & & & 1+24 & Else & Slope & Int. & + & - & Ini & Fin & R & L & \\
\midrule
BTC & Bear & 2021/12/1--2023/1/1 & $<$1 & -3.1 & 2.9 & -0.16 & -1.2 & 0.37 & -2.0 & 0.13 & 0.95 & 2.3 & 2.3 & -- \\
BTC & Bull & 2020/3/1--2021/4/1 & $<$1 & -7.9 & 2.8 & -0.12 & -1.1 & 0.73 & -3.4 & 0.26 & 0.13 & 2.4 & 2.2 & -- \\
ETH & Bear & 2021/12/1--2023/1/1 & $<$1 & -2.6 & 3.0 & -0.14 & -1.2 & 0.41 & -2.6 & 0.11 & 0.63 & 2.3 & 2.3 & -- \\
ETH & Bull & 2020/3/1--2021/4/1 & $<$1 & -7.5 & 2.9 & -0.12 & -1.1 & 0.6 & -3.7 & 0.22 & -0.23 & 2.5 & 2.4 & -- \\
\bottomrule
\end{tabular}
}
\caption{Statistics in Bull and Bear markets}
\label{table:regimes}
\end{table}

Other stylized facts exhibit notable regime-dependent behavior. Return autocorrelation patterns differ significantly from the full sample statistics (Table \ref{tab:asset_table}). During the bear market, both assets show insignificant autocorrelation at three standard errors, with "Avg ACF" values generally within $\pm3$ (BTC slightly lower at $-3.1$). However, the first and 24th-hour return autocorrelations fall well below $3$ in the bear market period. This pattern requires careful interpretation, as it might reflect market evolution rather than regime effects. Indeed, the more recent bear market period would naturally exhibit greater efficiency, and Section \ref{sec:half} suggests this evolutionary interpretation.

The leverage effect presents distinct patterns across regimes. Table \ref{table:regimes} shows negative cross-correlation (column "Ave Lev,-") between current returns and future absolute returns in both periods, but with varying intensity. The effect is weaker during the bear market (values above $-3$) compared to the bull market (values below $-3$: $-3.4$ for BTC, $-3.7$ for ETH). This asymmetry aligns with the characteristic volatility patterns of these regimes: the high volatility of bear markets may dampen the leverage effect, while the lower volatility during bull markets may amplify it.

Perhaps the most striking finding emerges from the Time Reversal Asymmetry (TRA) analysis. Table \ref{table:regimes} reveals a clear dichotomy: the TRA effect is notably absent during the bull market period but manifests strongly during the bear market phase.

\subsubsection{Halving Event Analysis}\label{sec:half}
To further examine the temporal stability of stylized facts, we analyze ETH and BTC across a significant structural break: the May 2020 Bitcoin halving event, when mining rewards decreased from 12.5 to 6.25 BTC. We select these tokens due to their extensive historical data and split our analysis into pre- and post-May 2020 periods. Our dataset encompasses this single halving event.

\begin{table}[ht]
\centering
\small
\resizebox{\textwidth}{!}{
\begin{tabular}{l c c r *{4}{c} *{2}{r} *{2}{c} *{2}{c} c}
\toprule
\multirow{2}{*}{Token} & \multirow{2}{*}{Halving} & \multirow{2}{*}{History} & \multirow{2}{*}{Zeros(\%)} & 
\multicolumn{2}{c}{Avg ACF} & \multicolumn{2}{c}{ACF Abs(R)} & \multicolumn{2}{c}{Avg Lev} & \multicolumn{2}{c}{TRA} & \multicolumn{2}{c}{CDF Tail} & \multirow{2}{*}{JB} \\
\cmidrule(lr){5-6} \cmidrule(lr){7-8} \cmidrule(lr){9-10} \cmidrule(lr){11-12} \cmidrule(lr){13-14}
& & & & 1+24 & Else & Slope & Int. & + & - & Ini & Fin & R & L & \\
\midrule
BTC & Before & 2017/8/17--2020/4/30 & $<$1 & -9.0 & 1.7 & -0.19 & -1.2 & 0.7 & -4.2 & 0.28 & 0.56 & 2.4 & 2.4 & -- \\
BTC & After & 2020/5/1--2023/10/19 & $<$1 & -3.6 & 1.9 & -0.16 & -1.1 & 1.3 & -3.0 & 0.13 & 0.38 & 2.7 & 2.6 & -- \\
ETH & Before & 2017/8/17--2020/4/30 & $<$1 & -8.5 & 1.7 & -0.29 & -1.2 & 0.9 & -4.4 & 0.29 & 0.41 & 2.5 & 2.4 & -- \\
ETH & After & 2020/5/1--2023/10/19 & $<$1 & -2.9 & 2.0 & -0.14 & -1.1 & 1.6 & -4.1 & 0.09 & 0.05 & 2.8 & 2.5 & -- \\
\bottomrule
\end{tabular}
}
\caption{Statistics before and after May 2020}
\label{table:half}
\end{table}

Table \ref{table:half} reveals several persistent statistical properties across this structural break. The unconditional probability density maintains heavy tails with tail coefficients showing minimal temporal variation. Similarly, the leverage effect remains consistently present, with comparable magnitudes in both periods.

Time Reversal Asymmetry exhibits asset-specific temporal variation. While BTC maintains significant TRA effects throughout both periods, ETH shows a marked change, with TRA disappearing in the post-2020 period. This finding aligns with our earlier observations in Table \ref{tab:asset_table} and Table \ref{table:regimes}, reinforcing that TRA's presence depends on both the specific token and the time period under consideration.

Volatility clustering persists across both periods but shows distinct characteristics. The autocorrelation function slopes indicate stronger persistence before May 2020 ($-0.19$ for BTC, $-0.29$ for ETH) compared to the post-halving period ($-0.16$ for BTC, $-0.14$ for ETH). This pattern aligns with our findings in Section \ref{sec:reg} regarding the bull and bear market regimes, which fall within the second half of our sample. Moreover, when comparing these results with the full sample statistics in Table \ref{tab:asset_table}, we observe that the first half of the data (pre-May 2020) dominates the overall sample characteristics.

The log returns' autocorrelation provides particularly instructive insights about market evolution. This metric, often considered the most fundamental measure of market efficiency, shows substantially stronger negative first-lag autocorrelation in the first half of the sample. This temporal pattern suggests a systematic increase in market efficiency following the halving event.

\subsection{Cross-Exchange Arbitrage and Price Synchronization}
The persistence of trading in seemingly less efficient pools, such as the 30 basis point (bp) pools, can be attributed to a crucial feature of modern DeFi functionality: cross-exchange arbitrage. This phenomenon is pivotal in maintaining market efficiency and price consistency across various trading venues.

Cross-exchange arbitrage occurs when the price of a token (e.g., ETH) on a 30 bp pool is more favorable than on other exchanges, even after accounting for higher transaction costs. This price discrepancy incentivizes arbitrageurs to trade across different exchanges and pools, capitalizing on the best net price after transaction costs. \footnote{Such arbitrage activities contribute significantly to the daily trading volume in cryptocurrency markets.} These arbitrage trades maintain price consistency across the diverse cryptocurrency exchange landscape.\footnote{There are more than 500 exchanges, according to https://coinmarketcap.com/.}

Our analysis reveals no significant differences in stylized facts between major centralized exchanges like Coinbase and Binance when examining the same historical period. These platforms' prices are very close, leading to consistent stylized facts (detailed in Appendix \ref{sec:cb}). The comparison between centralized exchanges (CEX) like Coinbase or Binance and decentralized exchanges (DEX) like Uniswap (on Arbitrum or Mainnet) presents a more nuanced picture. While price differences can be more pronounced between CEX and DEX, the stylized facts remain qualitatively identical. We posit that the persistent arbitrage opportunities are responsible for this consistency in stylized facts across DEX and CEX platforms.

A more detailed exploration of the price process on DEX and its relationship with CEX is provided in Section \ref{sec:dex}. We show that price changes on CEX often drive corresponding changes on DEX in Section \ref{sec:dex}. The resulting synchronization in price movements across these platforms contributes to the similarity in observed stylized facts.

This interconnectedness of price dynamics across various exchange types underscores the sophisticated nature of the cryptocurrency market ecosystem. It highlights how arbitrage activities ensure price efficiency and contribute to the consistent manifestation of stylized facts across diverse trading environments, from traditional centralized exchanges to innovative decentralized platforms.

\subsection{Cluster Analysis of Tokens Based on Stylized Facts}
Expanding our analysis beyond ETH on Binance and Uniswap, we turn to tokens trading on Coinbase. To comprehensively compare tokens based on their stylized facts, we employed a cluster analysis approach using the statistics presented in Table \ref{tab:asset_table}.

We construct a distance matrix using the Euclidean distance between rows of Table \ref{tab:asset_table}. The process involved normalizing each column after "Zero(\%)" by subtracting its mean and dividing by its standard deviation. We then calculate the Euclidean distance (square root of the sum of squared differences) between each pair of rows (tokens). This resulted in a square symmetric matrix with zeros on the diagonal, representing the distances between tokens based on their stylized fact statistics.

\begin{figure}[ht]
\centering
\includegraphics[width=1\linewidth]{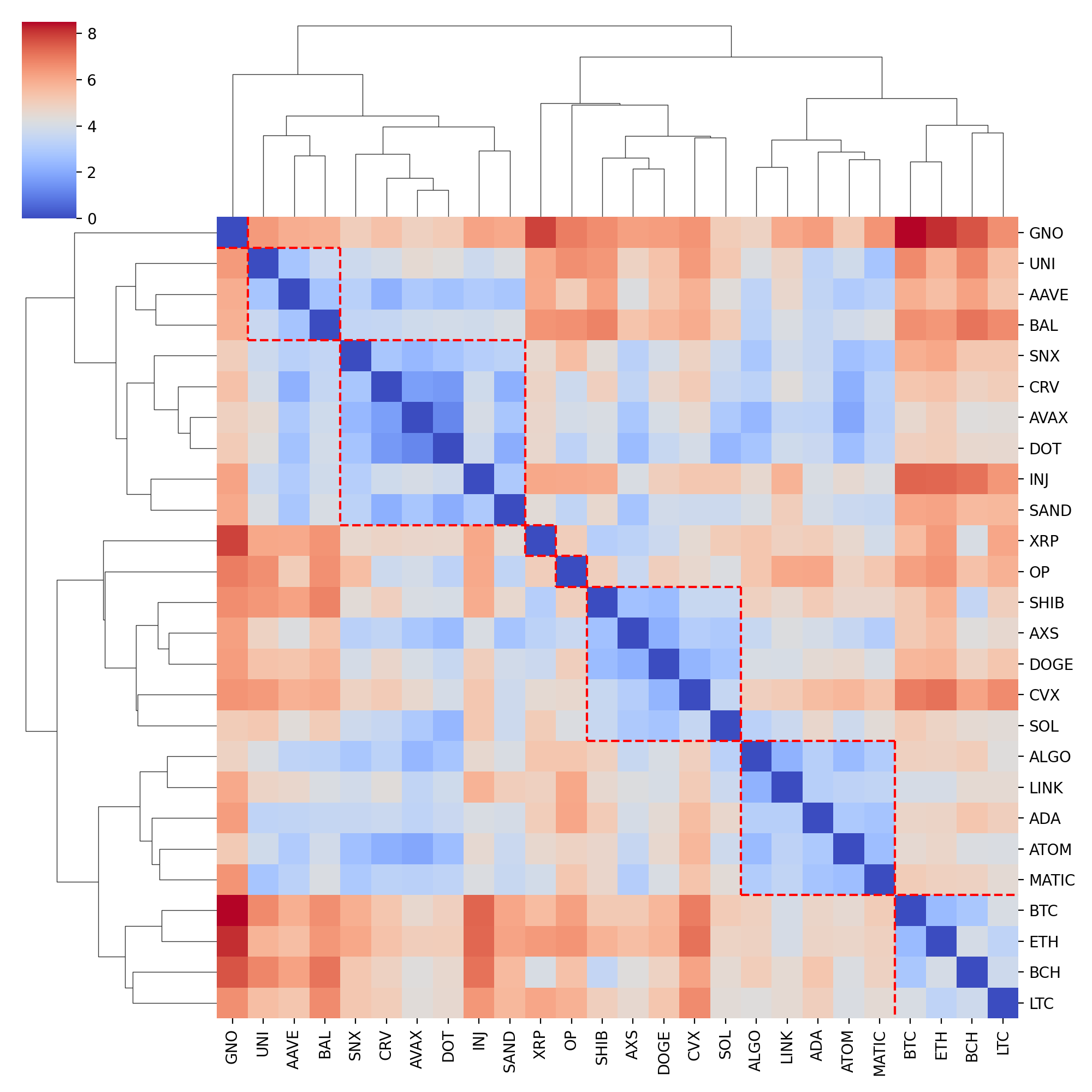}
\caption{Distance matrix between the vectors of stylized fact statistics in Table \ref{tab:asset_table} ordered using hierarchical clustering (method complete). We have highlighted 8 possible clusters. Tokens within the clusters have similar stylized facts.}
\label{fig:distmat}
\end{figure}

Figure \ref{fig:distmat} presents the resulting distance matrix, sorted using hierarchical clustering. We identified eight potential clusters, highlighted by red dashed boxes.

Key observations from the cluster analysis revealed several interesting patterns. GMO stands out as a unique cluster, possibly due to its liquidity characteristics, though this requires further investigation. Especially because CVX belongs to a larger cluster despite its high zero return fraction, challenging the assumption that liquidity is a primary factor in determining stylized facts or cluster formation.

The first cluster from the right comprises the largest and oldest tokens in the Web3 universe, predominantly PoW tokens (except for ETH, recently converted to PoS). This clustering suggests that longevity and market share significantly influence stylized facts. Another notable cluster includes two major spot market DEXs (UNI and BAL) and AAVE, a lending/borrowing platform. The absence of CRV (another large spot DEX) from this group can be attributed to its focus on stablecoin trading, which contrasts with the general-purpose nature of UNI and BAL. AAVE's inclusion likely stems from its role in facilitating short-selling and leveraged positions, creating an interconnection with UNI and BAL.

The remaining clusters in Figure \ref{fig:distmat} show less clear delineations, combining tokens from various sectors as listed in Table \ref{tab:asset_table}. This mixture of sectors within clusters suggests that the relationships between tokens based on their stylized facts may transcend traditional sector classifications.

\subsection{Detailed Analysis of DEX vs CEX Dynamics} \label{sec:dex}
This section analyzes the price dynamics between decentralized exchange (DEX) pools (Arbitrum and Mainnet 30bp and 5bp WETH-USDC) and the centralized exchange (CEX) Binance ETH market, focusing on arbitrage opportunities and price discovery mechanisms between these platforms.

\subsubsection{Arbitrage Mechanics in DEX}
Arbitrage bots play a crucial role in DeFi ecosystems by exploiting price discrepancies between pools and centralized exchanges. Our analysis aims to identify these arbitrage activities in empirical data. In an idealized scenario where arbitrageurs are the sole traders in a pool, they trade until the pool price reaches an optimal target value, beyond which the price difference would not yield a profit after accounting for trading fees.

Let $Z$ represent the pool price, $S$ the CEX price, and $\gamma$ the fee constant (where $\gamma = 1 - 0.3\%$ for 30bp pools and $1 - 0.05\%$ for 5bp pools). The optimal target pool price after arbitrage is theoretically:
\begin{equation}
Z_\text{opt} =
\begin{cases}
\gamma^{-1} S & \text{if } Z > \gamma^{-1} S \\
Z             & \text{if } Z \in  [\gamma S, \gamma^{-1} S]  \\
\gamma S      & \text{if } Z < \gamma S
\end{cases}
\label{eq:dex}
\end{equation}

\noindent The interval $[\gamma S, \gamma^{-1} S]$ is termed the "no-arbitrage region," where arbitrageurs cannot profit due to trading fees. We refer the interested reader to \cite{tassywhite,10.1145/3419614.3423251, 2023arXiv230514604M} for further details.

\subsubsection{Empirical Observations}
\begin{figure}[ht]
\centering
\includegraphics[width=1\linewidth]{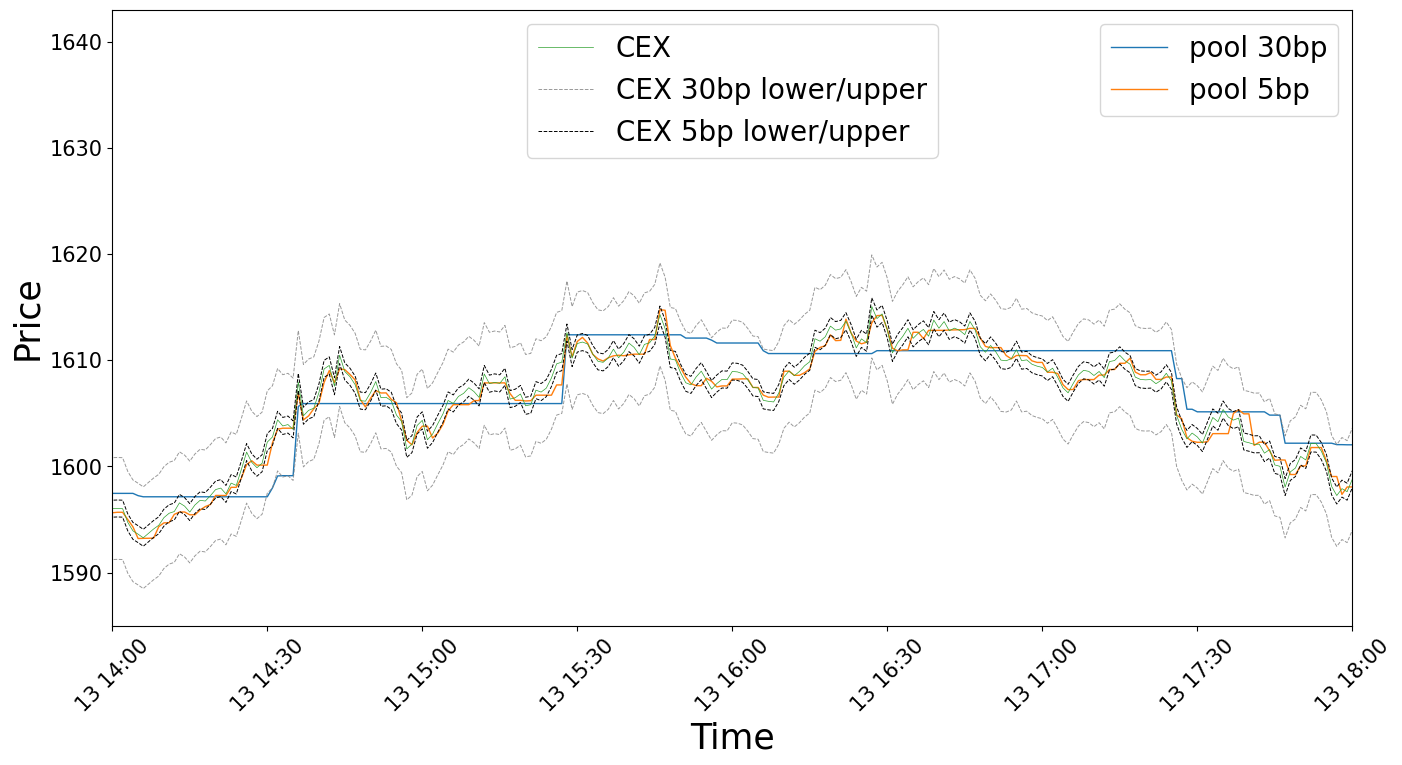}
\caption{Price time series (1-minute timescale) from 2023/9/13 14:00:00 to 2023/9/13 18:00:00. Dashed lines indicate the upper and lower boundaries of the no-arbitrage region.}
\label{fig:pool_CEX_timeseries}
\end{figure}

Figure \ref{fig:pool_CEX_timeseries} illustrates the price time series (1-minute intervals) for 2023/9/13 14:00:00 to 2023/9/13 18:00:00, including the no-arbitrage region boundaries. The 30 bp pool price remains consistent within the no-arbitrage region, only moving when it touches the boundaries. The same behavior can be observed for the 5 bp pool, but since the 5 bp is tighter, the price jumps more often than in the 30 bp pool.

Figure \ref{fig:pool_CEX_timeseries} suggests that CEX leads DEX in price discovery. The price discovery appears to happen on CEX, and only when it is profitable does the price on a DEX change to match the new CEX price \cite{10.1145/3419614.3423251, 2023arXiv230514604M, TungYen2024}. We can further formalize this observation by examining the correlation between lagged ETH returns on a CEX and on a DEX.

\subsubsection{Price Discovery and Lead-Lag Relationships}
Specifically, we analyze the cross-correlation of returns on a 10-minute scale. Let $R_B$ denote Binance (CEX) ETH returns and $R_p$ the Uniswap (DEX) ETH pool returns. For a given lag $k \in \mathbb{Z}$, the cross-correlation is computed as:
\begin{equation}
\frac{ \mathbb{E} \left[R_B(t) R_p(t+k)\right]}{\sigma_{{R_B}} \sigma_{{R_p}}}
\end{equation}

\begin{figure}[ht]
\centering
\includegraphics[width=1\linewidth]{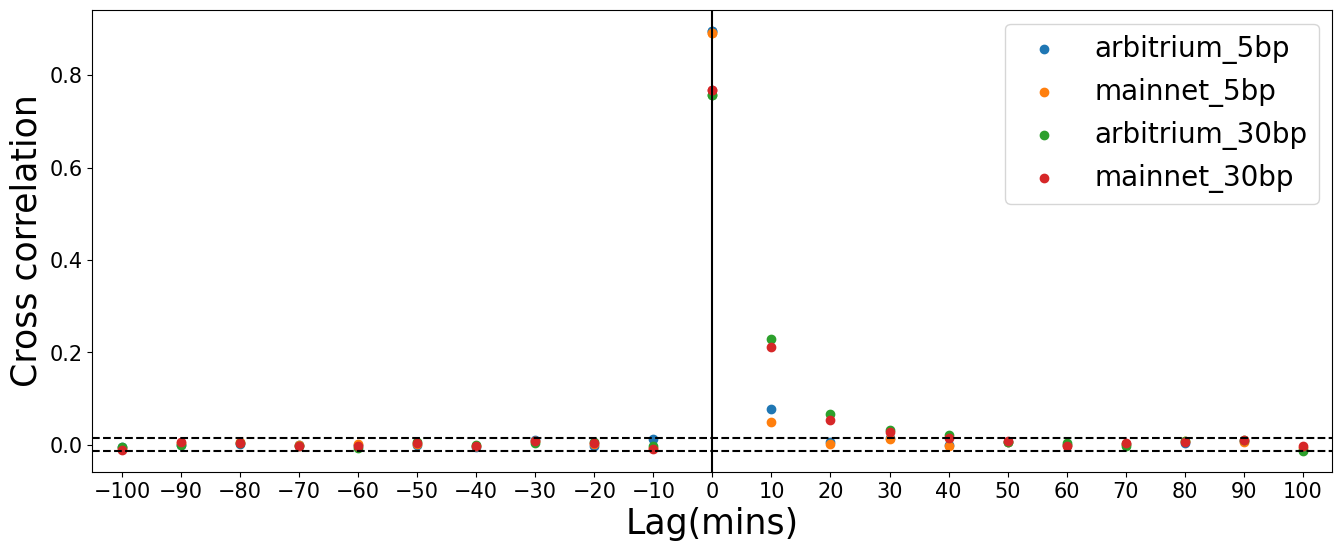}
\caption{Cross-correlation between DEX pool and CEX returns, the error bars are $\pm 2$ standard errors.}
\label{fig:pool_CEX_cc}
\end{figure}

Figure \ref{fig:pool_CEX_cc} displays this cross-correlation. The result suggests that CEX price affects the pool price from now to twenty minutes in the future but not vice versa, providing evidence that the pool price is pushed by CEX.

The observed patterns in price movements and cross-correlations provide several insights into the dynamics between DEX and CEX. The 30bp pool's strict adherence to the no-arbitrage region suggests highly efficient arbitrage mechanisms, likely due to the higher fees attracting more sophisticated traders. In contrast, the 5bp pool's occasional excursions outside the no-arbitrage region indicate a more diverse trader base, potentially including more retail or less latency-sensitive participants.

The cross-correlation analysis reveals important lead-lag relationships between DEX and CEX, providing clear evidence that CEX leads in price discovery. This finding has significant implications for understanding information flow in cryptocurrency markets and could inform trading strategies that exploit these temporal dynamics.

\subsubsection{Implications of Fee Structures}
Furthermore, the differences in behavior between 30bp and 5bp pools highlight how fee structures can significantly impact trading dynamics and market efficiency. The higher fees in the 30bp pool appear to create a more stable price environment, while the lower fees in the 5bp pool lead to more frequent price adjustments, potentially offering more opportunities for nimble traders \cite{2023arXiv230514604M}.

Lastly, the persistence of price discrepancies, particularly in the 5bp pool, suggests ongoing arbitrage opportunities for sophisticated traders. This observation underscores the complexity of the cryptocurrency market ecosystem, where different trading venues and fee structures create a rich landscape of opportunities and challenges for market participants.

\section{Acknowledgement}
The authors gratefully acknowledge Wun-Cing Liou and Min-Ren Guan for their essential contributions to blockchain data collection and Wei-Lun Tao for his meticulous work in cleaning traditional finance data.

ACS acknowledges the generous support and hospitality of the National Center for Theoretical Sciences, Mathematics Division, at the National Taiwan University, and the Department of Mathematics at the National Tsing Hua University during the summer of 2024.

This research was supported in part by NSTC grant 111-2115-M-007-014-MY3 (S.N. Tung).

\section*{Appendix}

\section*{Appendix 1. Tail Behavior Analysis} \label{app:tail} 
Financial time series exhibit ``fat tails'' characterized by power-law decay, particularly at short time scales. We investigate this phenomenon by analyzing the Cumulative Distribution Function (CDF).

We normalize return data and compute the empirical CDF:
\begin{equation}
\hat{F}(x) \equiv \frac{1}{n} \sum^{n}_{i=1} \mathbf{1}_{ X_i \leq x}
\end{equation}

To visualize tail behavior, we plot the logarithm of the following:
\begin{equation} \label{eq:ccdf}
\begin{cases}
1- \log{\hat{F}(x)} & \text{if}\ x \geq 0
\quad (\text{denoted by} \ \square) \\
\log{\hat{F}(-x)} & \text{if}\ x < 0
\quad (\text{denoted by} \  \blacktriangle)
\end{cases}
\end{equation}

\begin{figure}[ht]
\centering
\includegraphics[width=0.8\linewidth]{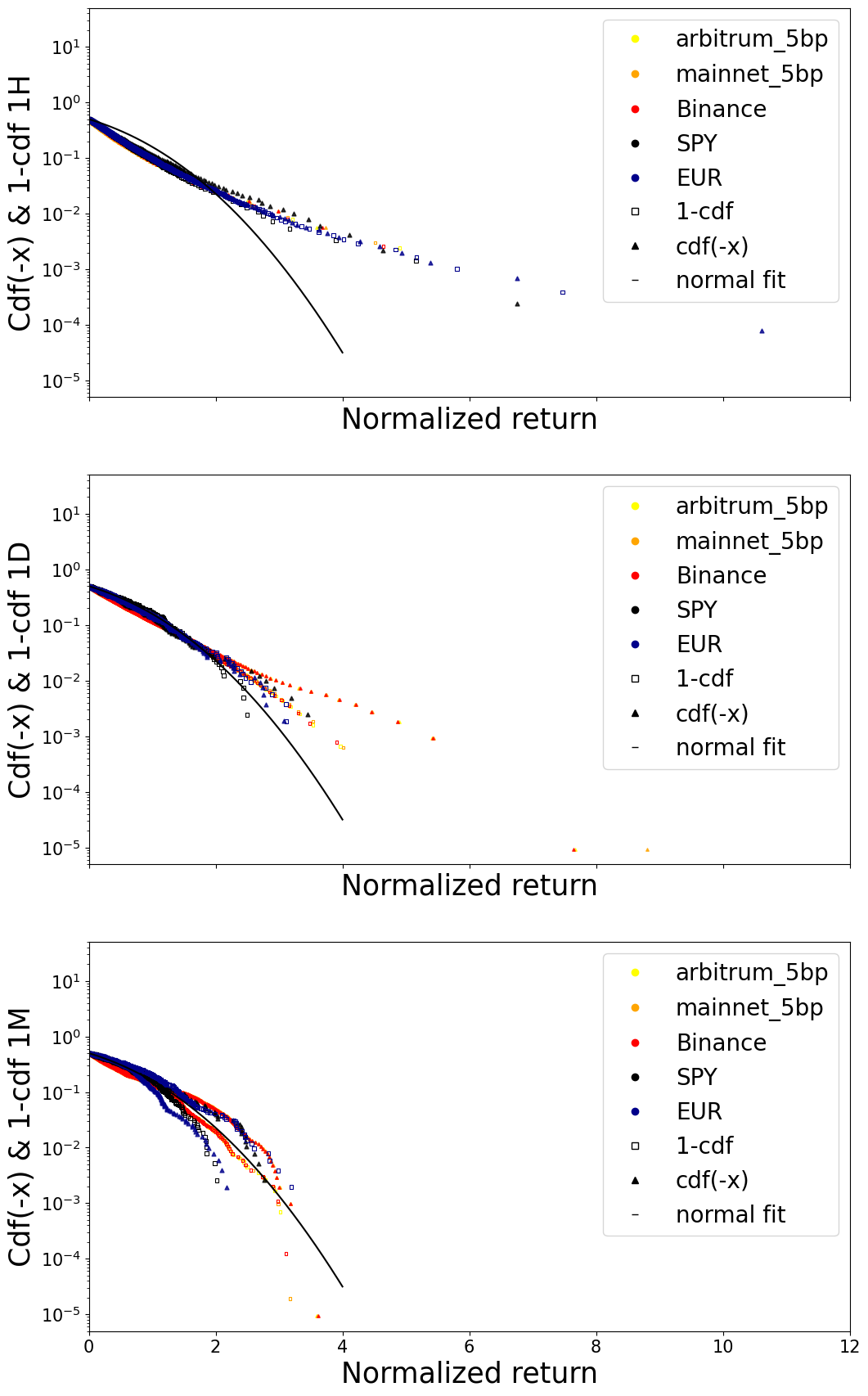}
\caption{Cumulative Distribution Function in log-linear scale using Equation~(\ref{eq:ccdf}) for pool, CEX ETH, and traditional finance returns at one-hour (1H), one-day (1D), and one-month (1M) time scales. Assets include Arbitrum 5bp WETH-USDC, Mainnet 5bp WETH-USDC, Binance ETH, SPY, and EUR. CDF deviates from the standard normal distribution at short return intervals.}
\label{fig:pool_CEX_cdf}
\end{figure}

\begin{figure}[ht]
\centering
\includegraphics[width=1\linewidth]{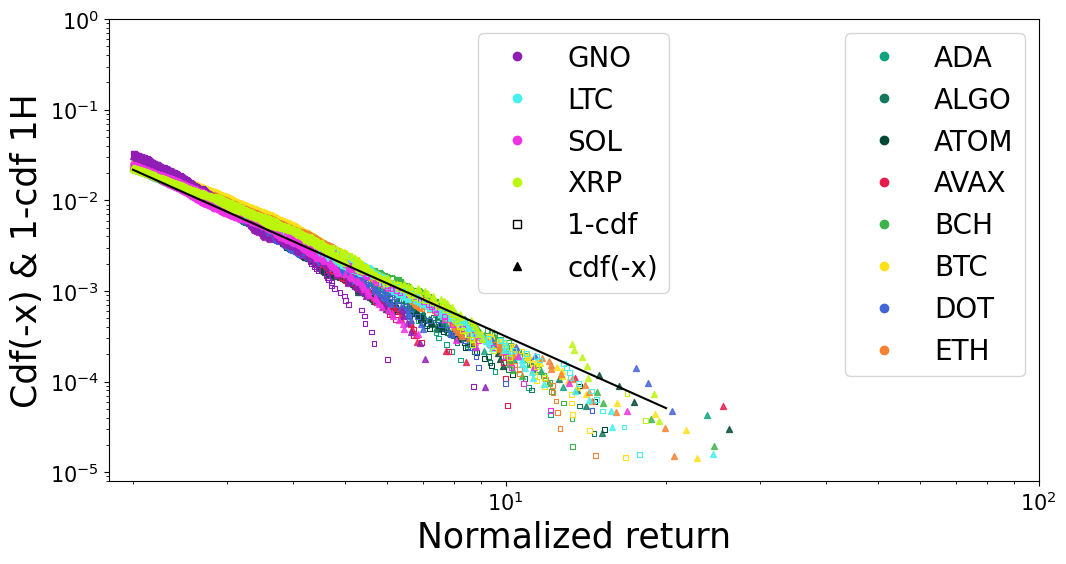}
\caption{Log-log plot of CDF tails using Equation~(\ref{eq:ccdf}) for L1 tokens' one-hour normalized returns. A power law fit $\sim |x|^{-\alpha}$ with $\alpha = 2.63$ is overlaid for comparison.}
\label{fig:L1_cdf_log_log}
\end{figure}

\begin{figure}[ht]
\centering
\includegraphics[width=1\linewidth]{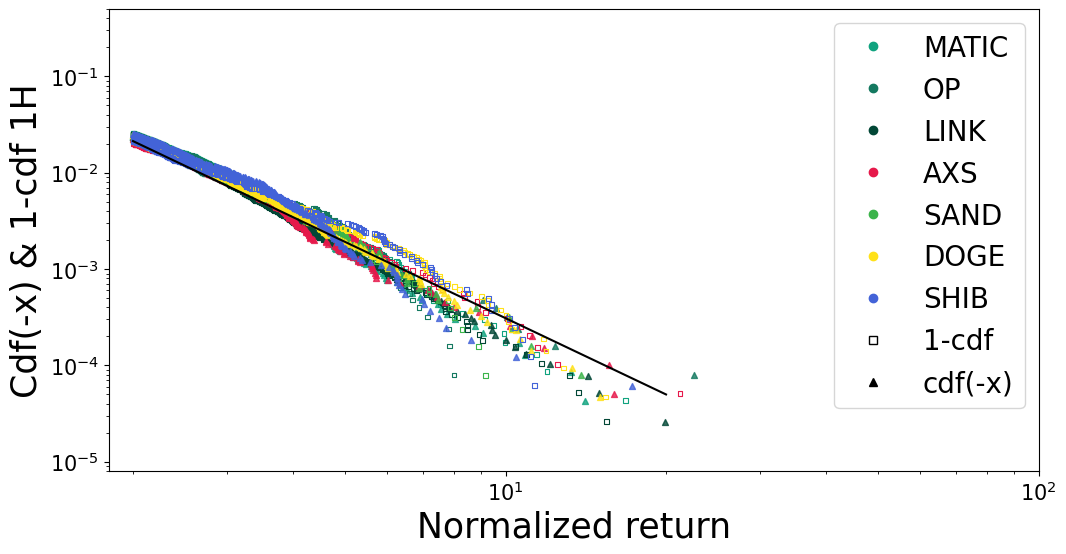}
\caption{Log-log plot of CDF tails using Equation~\eqref{eq:ccdf} for miscellaneous tokens' one-hour normalized returns. A power law fit $\sim |x|^{-\alpha}$ with $\alpha= 2.63$ is included for reference.}
\label{fig:L2+others_cdf_log_log.png}
\end{figure}

\begin{figure}[ht]
\centering
\includegraphics[width=1\linewidth]{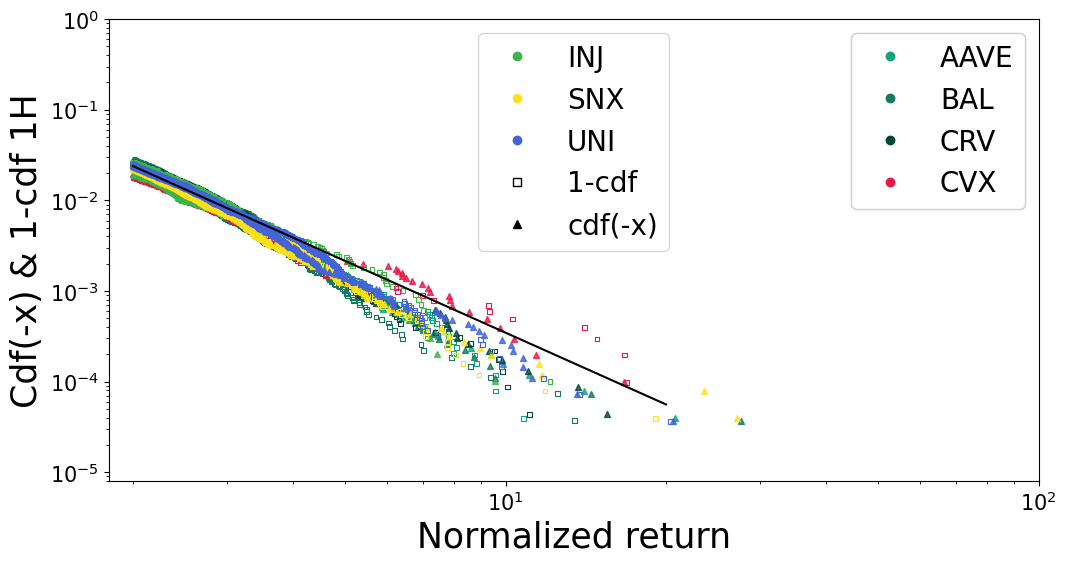}
\caption{Log-log plot of CDF tails using Equation~\eqref{eq:ccdf} for DeFi tokens' one-hour normalized returns. A power law fit $\sim |x|^{-\alpha}$ with $\alpha = 2.63$ is shown for comparison.}
\label{fig:defi_cdf_log_log}
\end{figure}

We flip the $x<0$ part to the positive x-axis, enabling simultaneous examination of both tails. Figures~\ref{fig:pool_CEX_cdf}, \ref{fig:L1_cdf_log_log}, \ref{fig:L2+others_cdf_log_log.png}, and \ref{fig:defi_cdf_log_log} display Equation~\eqref{eq:ccdf} for various assets and timescales.

Figure \ref{fig:pool_CEX_cdf} shows the power-law tails for one-hour returns and the progression to a Gaussian distribution of the CDF presented as in Equation \ref{eq:ccdf}. This is equivalent to Figure \ref{fig:pool_CEX_pdf} for the PDF but with an increased emphasis on the tails. 

Figures \ref{fig:L1_cdf_log_log}, \ref{fig:L2+others_cdf_log_log.png}, and \ref{fig:defi_cdf_log_log} extend the analysis to all other tokens in Table \ref{tab:asset_table} by presenting the one-hour returns CDF (Equation \ref{eq:ccdf}). The tails of the CDF are power law with a coefficient of $\alpha \sim 2.63$, similar to the value in the main text.

\section*{Appendix 2. Autocorrelation details} \label{app:acfR}
To enable comparison between traditional assets and Web3 tokens, we implement a custom autocorrelation function (ACF) algorithm. This approach addresses the discrepancy in trading hours between traditional assets (e.g., SPY trades for 6.5 hours) and continuously traded Web3 tokens.

We compute the ACF at lag $k$ for SPY using the following equation:
\begin{equation} \label{eq:acf_non-24/7}
\frac{  \frac{1}{\sqrt{N}}\displaystyle
\sum_{(R_t,R_{t+k}) \in I_k} (R_t - \hat{\mu}_1) (R_{t+k} - \hat{\mu}_2 )}{\hat{\sigma}_1 \hat{\sigma}_2 }
\end{equation}

\noindent where $R_t$ is the return at time $t$, $I_k$ is the set of return pairs for lag $k$, $N$ is the size of $I_k$, and $\hat{\mu}_1, \hat{\mu}_2, \hat{\sigma}_1, \hat{\sigma}_2$ are empirical means and standard deviations of the respective return sets.

Equation~\eqref{eq:acf_non-24/7} calculates ACF for every hour in a day, resulting in "gap" periods with no pairs of hours for SPY (e.g., at a lag of 12 hours).

\begin{figure}[ht]
\centering
\includegraphics[width=1\linewidth]{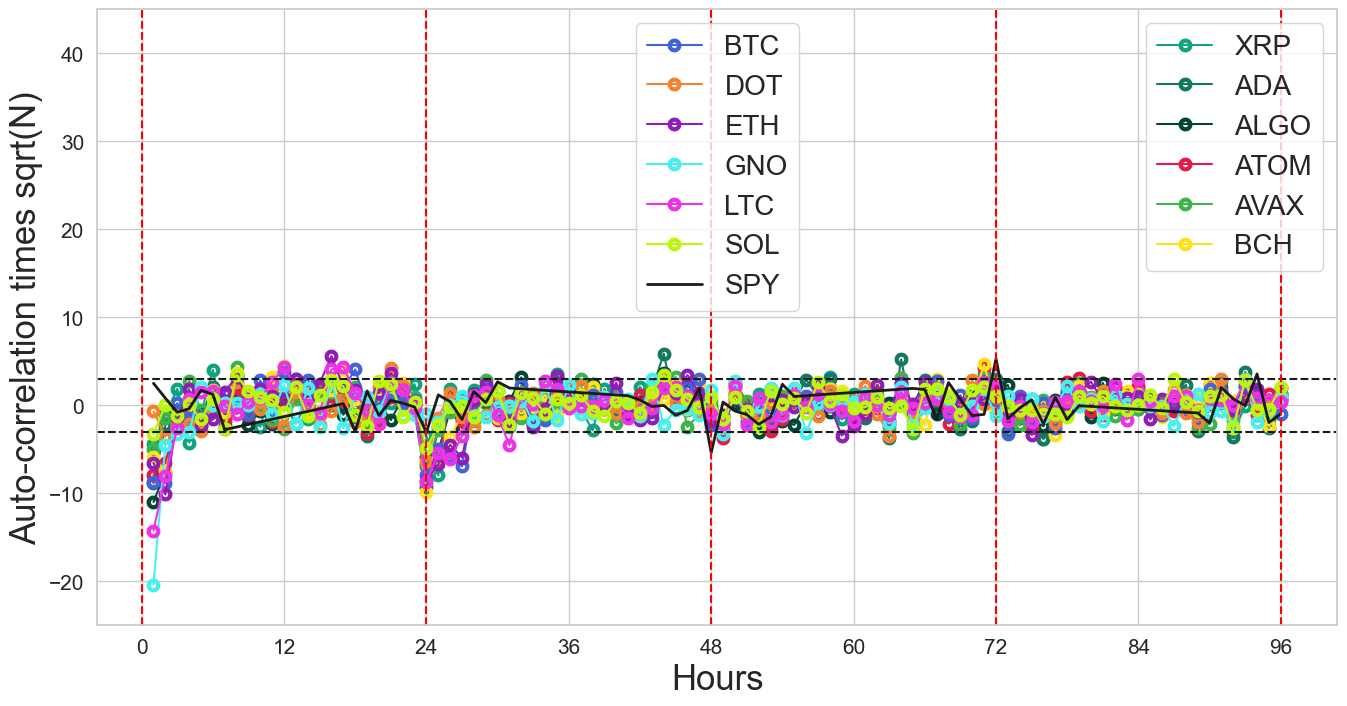}
\caption{Autocorrelation function of L1 tokens' one-hour returns.}
\label{fig:L1_acf_R}
\end{figure}

\begin{figure}[ht]
\centering
\includegraphics[width=1\linewidth]{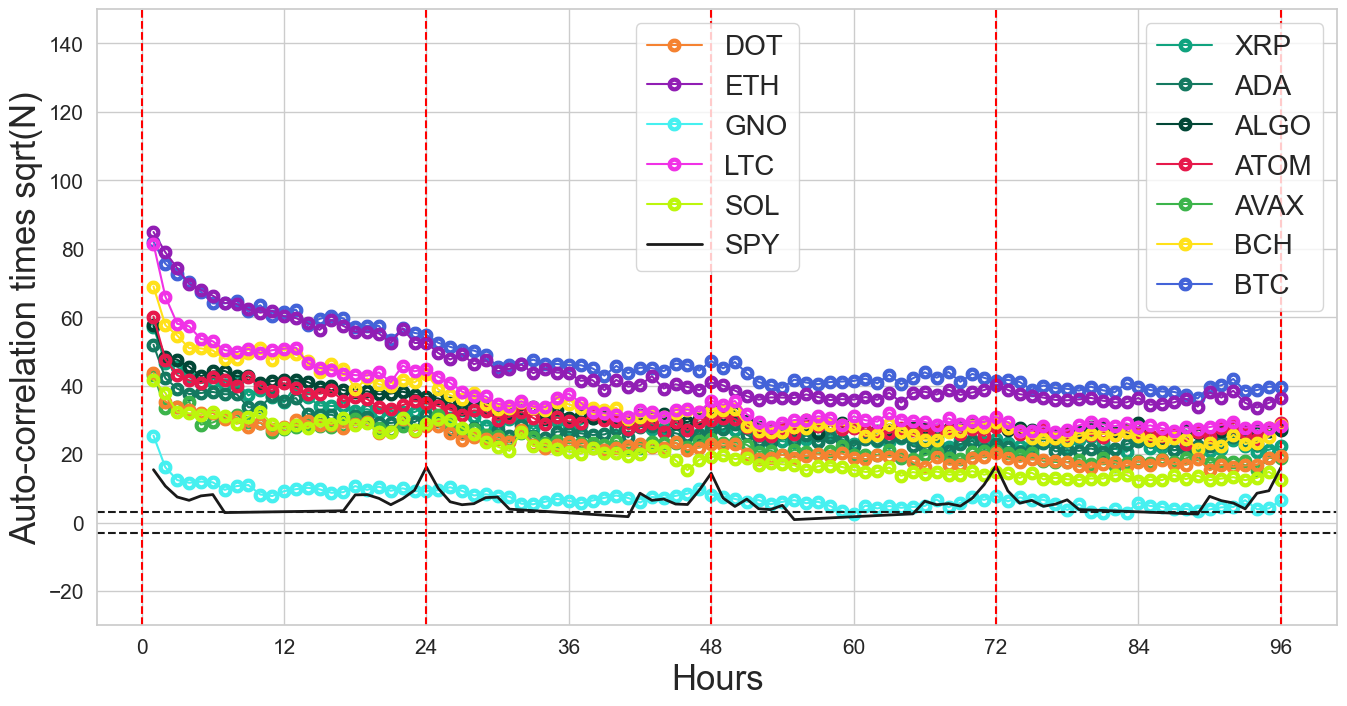}
\caption{Autocorrelation function of L1 tokens' one-hour absolute returns.}
\label{fig:L1_acf_absR}
\end{figure}

\begin{figure}[ht]
\centering
\includegraphics[width=1\linewidth]{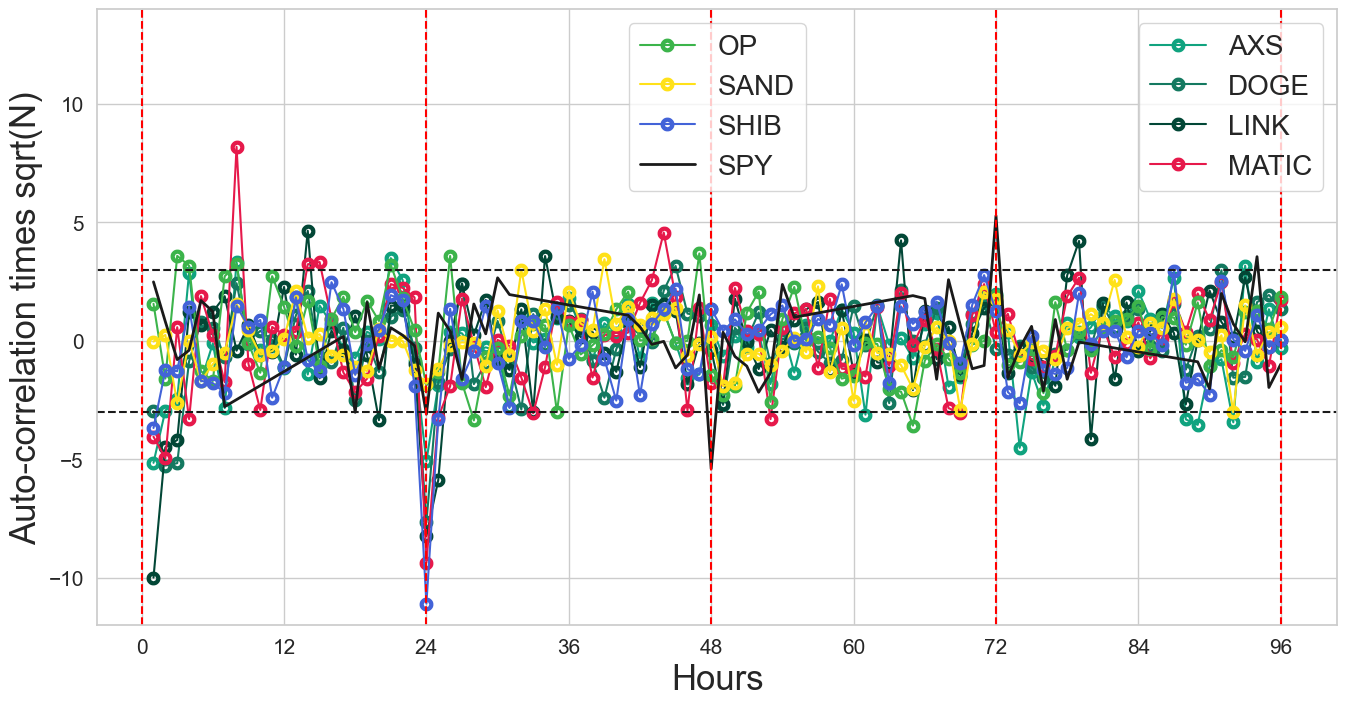}
\caption{Autocorrelation function of miscellaneous tokens' one-hour returns.}
\label{fig:L2+others_acf_R}
\end{figure}

\begin{figure}[ht]
\centering
\includegraphics[width=1\linewidth]{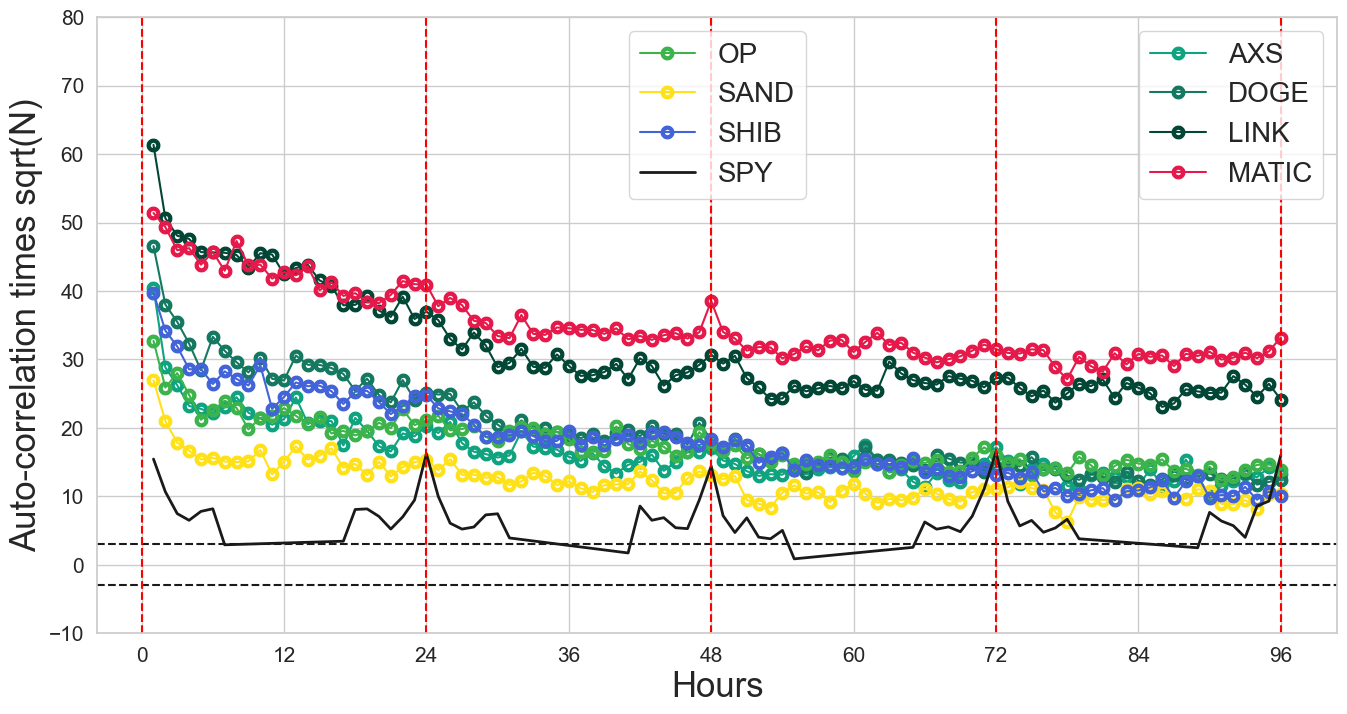}
\caption{Autocorrelation function of miscellaneous tokens' one-hour absolute returns.}
\label{fig:L2+others_acf_absR}
\end{figure}

\begin{figure}[ht]
\centering
\includegraphics[width=1\linewidth]{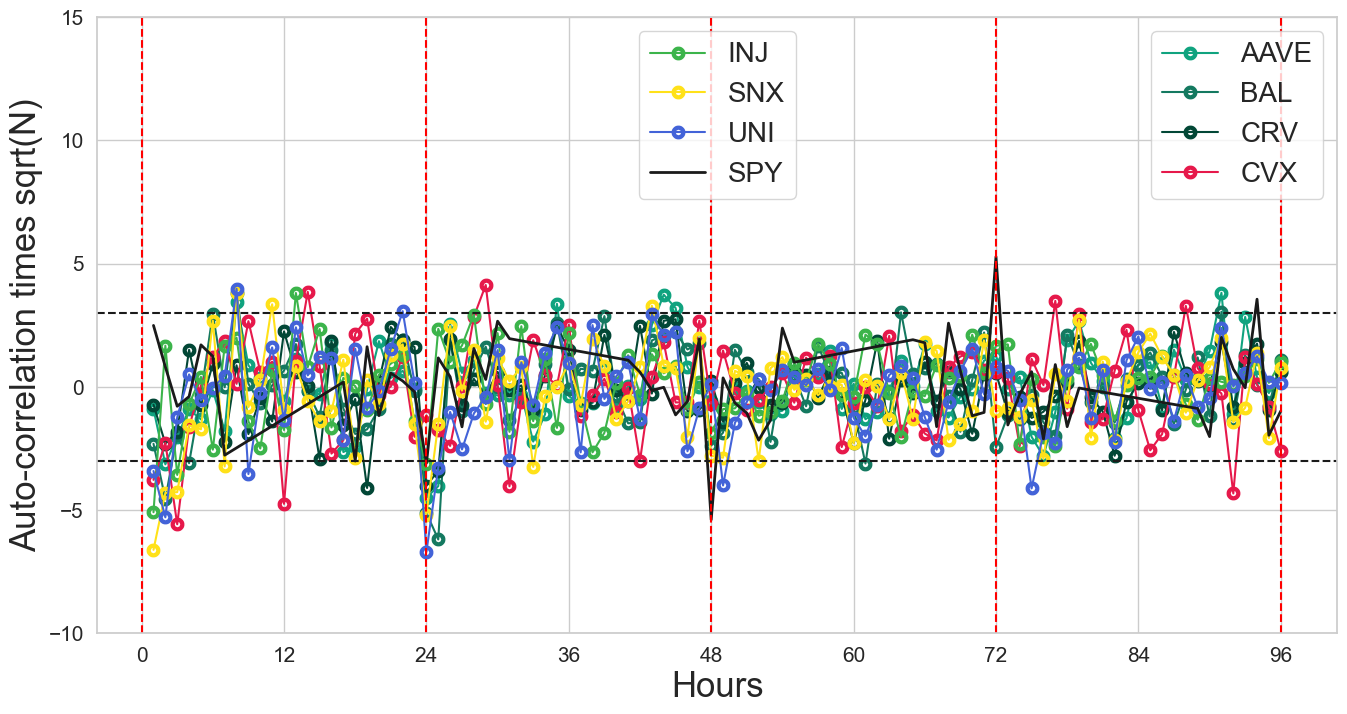}
\caption{Autocorrelation function of DeFi tokens' one-hour returns.}
\label{fig:defi_acf_R}
\end{figure}

\begin{figure}[ht]
\centering
\includegraphics[width=1\linewidth]{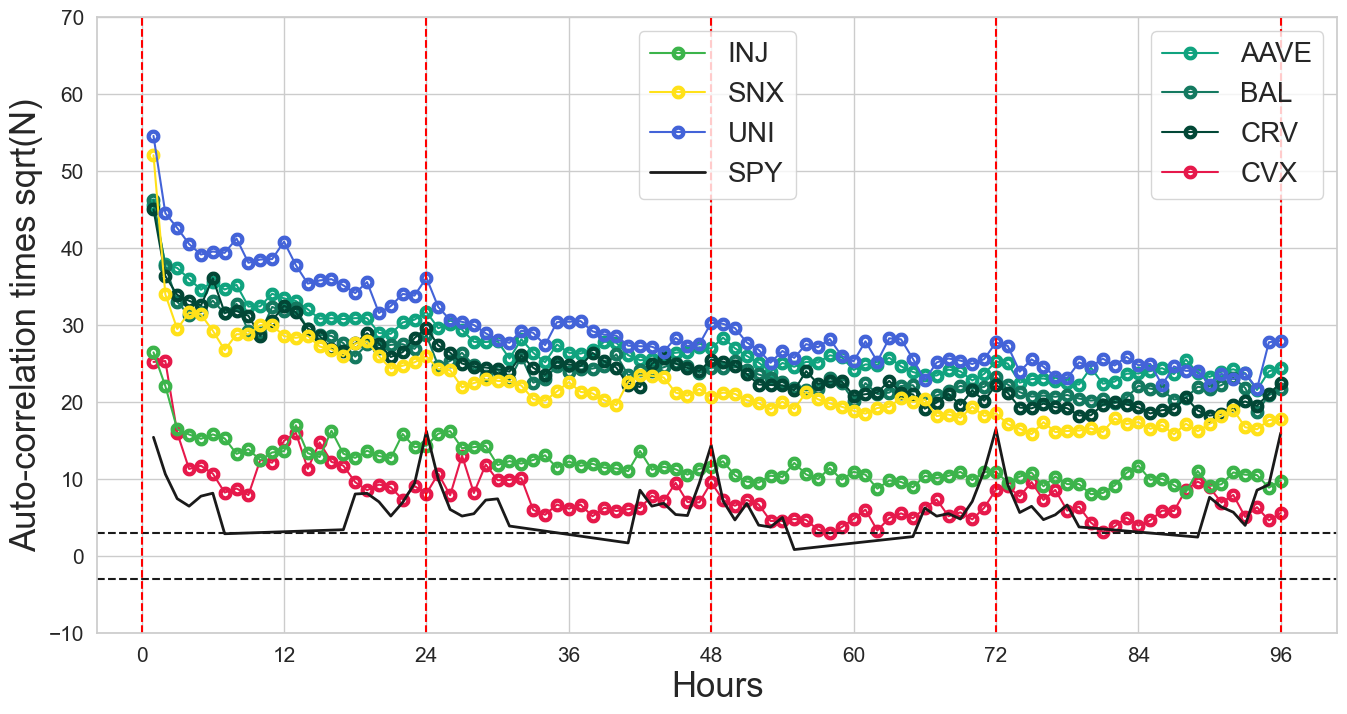}
\caption{Autocorrelation function of DeFi tokens' one-hour absolute returns.}
\label{fig:defi_acf_absR}
\end{figure}

Figures~\ref{fig:L1_acf_R}, \ref{fig:L1_acf_absR}, \ref{fig:L2+others_acf_R}, \ref{fig:L2+others_acf_absR}, \ref{fig:defi_acf_R}, and \ref{fig:defi_acf_absR} display autocorrelations of returns and absolute returns for tokens listed in Table~\ref{tab:asset_table}, using all available historical data. These figures support the main text findings: return autocorrelations are mostly noise except for specific lags, while absolute return autocorrelations are large and decay slowly, indicating volatility clustering.

\section*{Appendix 3. Random Zeros Replacement} \label{app:zero}
\begin{figure}[ht]
\centering
\includegraphics[width=1\linewidth]{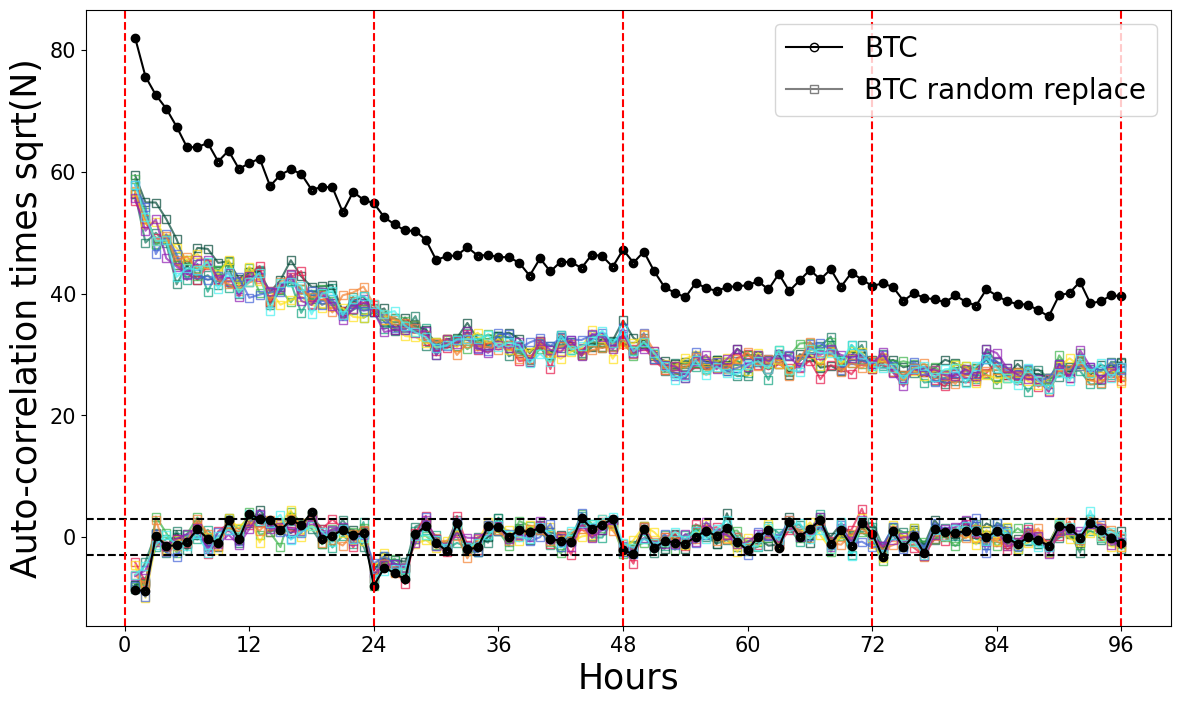}
\caption{Autocorrelation of one-hour absolute returns for Coinbase BTC (black marks) and the zeros-replaced samples with replacement rate $r = 0.24$ (colored marks)}
\label{fig:random_replacement_1}
\end{figure}

We investigate the impact of zero returns on the autocorrelation function (ACF) using Coinbase BTC data. We randomly replace 24\% of data points with zeros (matching GNO's zero-return proportion) in 10 modified datasets. Figure~\ref{fig:random_replacement_1} compares ACFs of returns and absolute returns between modified and original data.

Results show minimal impact on returns ACF but a consistent downward shift in absolute returns ACF for modified datasets. This suggests zero returns don't significantly affect price changes' linear dependence but can lead to underestimated volatility persistence, potentially explaining lower ACF levels in tokens with high zero-return proportions.

\section*{Appendix 4. Coinbase vs Binance Comparison} \label{sec:cb}
\begin{figure}[ht]
\centering
\includegraphics[width=1\linewidth]{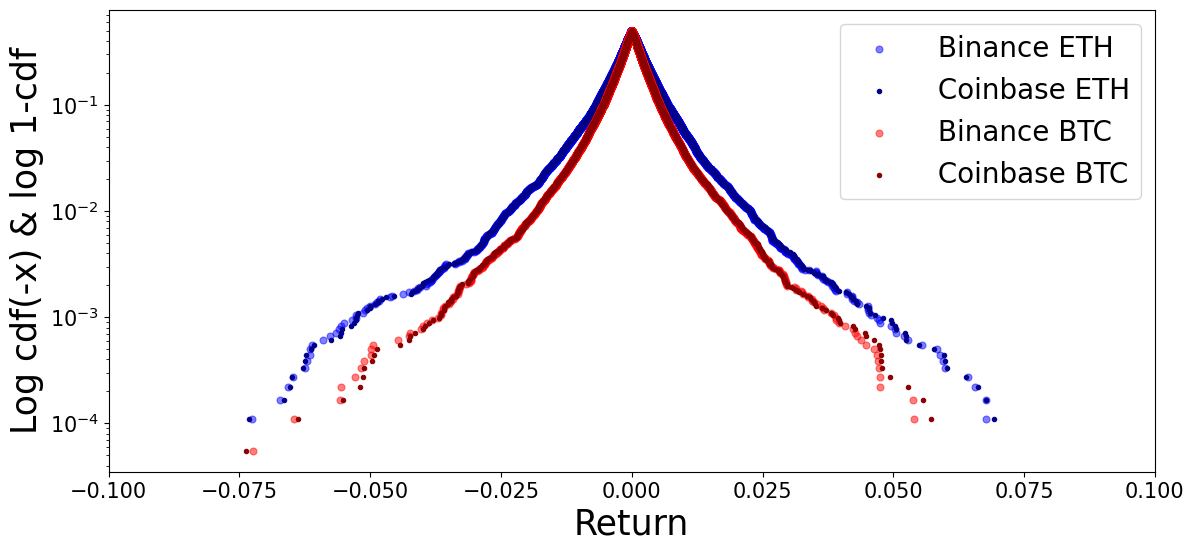}
\caption{CDF of Binance and Coinbase ETH, BTC returns from 2021/9/23 to 2023/10/18. We show a mountain plot: CDF(x) for x<0 and 1-CDF(x) for x>0.}
\label{fig:/CEX_compare_cdf_partial}
\end{figure}

\begin{figure}[ht]
\centering
\includegraphics[width=1\linewidth]{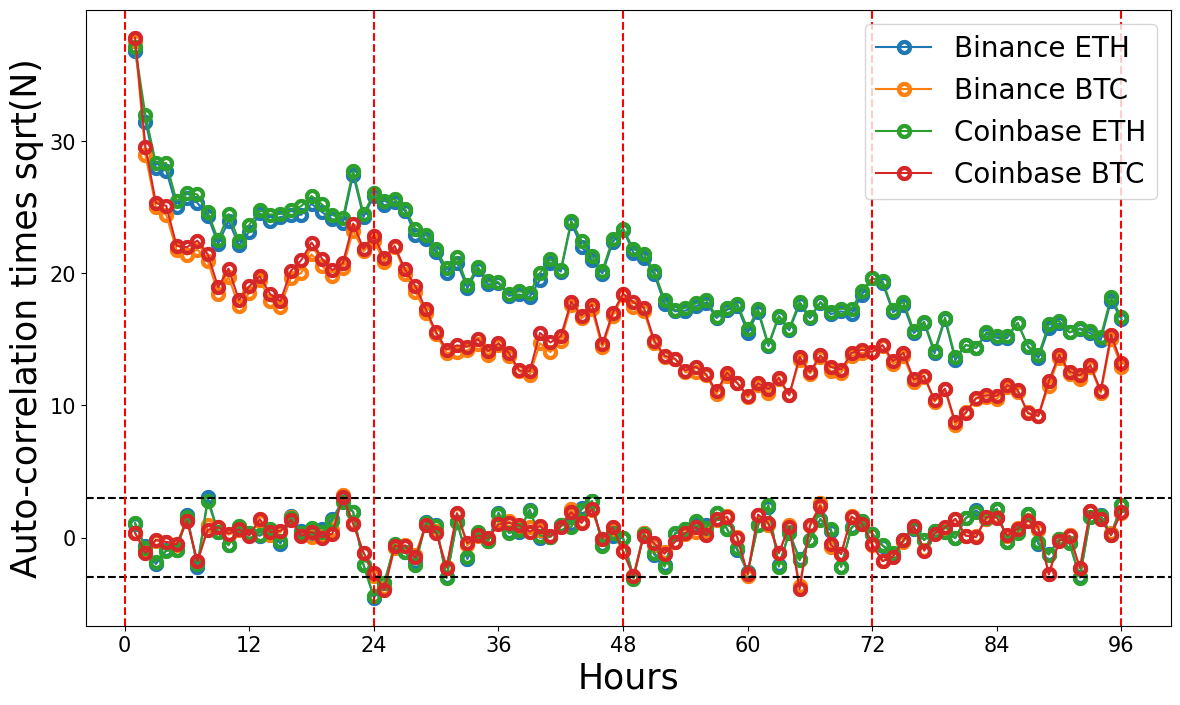}
\caption{Autocorrelation functions of Binance and Coinbase ETH, BTC return \& absolute return for data from 2021/9/23 to 2023/10/18. The curves for the same token overlap.}
\label{fig:/CEX_compare_acf_partial}
\end{figure}

We compare Coinbase and Binance data for ETH and BTC from September 23, 2021, to October 18, 2023. Figure~\ref{fig:/CEX_compare_cdf_partial} shows one-hour returns' cumulative distribution functions, revealing minimal differences except at extreme tails. Figure~\ref{fig:/CEX_compare_acf_partial} compares autocorrelation functions for returns and absolute returns, showing largely indistinguishable patterns.

These findings demonstrate negligible price differences between Binance and Coinbase during the study period, reinforcing the consistency of observed stylized facts in cryptocurrency markets.

\bibliographystyle{plain}
\bibliography{Reference}

\begin{thebibliography}{10}

\bibitem{10.1145/3419614.3423251}
Guillermo Angeris and Tarun Chitra.
\newblock Improved price oracles: Constant function market makers.
\newblock AFT '20, page 80–91, New York, NY, USA, 2020. Association for Computing Machinery.

\bibitem{Chakraborti2011}
Marco~Patriarca Anirban~Chakraborti, Ioane Muni~Toke and Frédéric Abergel.
\newblock Econophysics review: I. empirical facts.
\newblock {\em Quantitative Finance}, 11(7):991--1012, 2011.

\bibitem{Bouchaud_Potters_2003}
Jean-Philippe Bouchaud and Marc Potters.
\newblock {\em Theory of Financial Risk and Derivative Pricing: From Statistical Physics to Risk Management}.
\newblock Cambridge University Press, 2 edition, 2003.

\bibitem{EthereumWhitepaper}
Vitalik Buterin.
\newblock Ethereum whitepaper.
\newblock \url{https://ethereum.org/en/whitepaper/}, 2013.

\bibitem{Cont2001}
R.~Cont.
\newblock Empirical properties of asset returns: stylized facts and statistical issues.
\newblock {\em Quantitative Finance}, 1(2):223--236, 2001.

\bibitem{harvey2021defi}
Campbell~R. Harvey, Ashwin Ramachandran, and Joey Santoro.
\newblock {\em DeFi and the Future of Finance}.
\newblock John Wiley \& Sons, Hoboken, New Jersey, 2021.

\bibitem{HESTON2008418}
Steven~L. Heston and Ronnie Sadka.
\newblock Seasonality in the cross-section of stock returns.
\newblock {\em Journal of Financial Economics}, 87(2):418--445, 2008.

\bibitem{lipton2022blockchain}
Alexander Lipton and Adrien Treccani.
\newblock {\em Blockchain and Distributed Ledgers: Mathematics, Technology, and Economics}.
\newblock World Scientific, Singapore, 2022.

\bibitem{Mantegna_Stanley_1999}
Rosario~N. Mantegna and H.~Eugene Stanley.
\newblock {\em Introduction to Econophysics: Correlations and Complexity in Finance}.
\newblock Cambridge University Press, 1999.

\bibitem{2023arXiv230514604M}
Jason {Milionis}, Ciamac~C. {Moallemi}, and Tim {Roughgarden}.
\newblock {Automated Market Making and Arbitrage Profits in the Presence of Fees}.
\newblock {\em arXiv e-prints}, page arXiv:2305.14604, May 2023.

\bibitem{TungYen2024}
Joseph Najnudel, Shen-Ning Tung, Kazutoshi Yamazaki, and Ju-Yi Yen.
\newblock An arbitrage driven price dynamics of automated market makers in the presence of fees, 2024.

\bibitem{nakamoto2008bitcoin}
Satoshi Nakamoto.
\newblock Bitcoin: A peer-to-peer electronic cash system.
\newblock \url{https://bitcoin.org/bitcoin.pdf}, 2008.

\bibitem{Rose2020}
Radoš~Radoičić Omar El~Euch, Jim~Gatheral and Mathieu Rosenbaum.
\newblock The zumbach effect under rough heston.
\newblock {\em Quantitative Finance}, 20(2):235--241, 2020.

\bibitem{PLEROU2000374}
V~Plerou, P~Gopikrishnan, B~Rosenow, L.A.N Amaral, and H.E Stanley.
\newblock A random matrix theory approach to financial cross-correlations.
\newblock {\em Physica A: Statistical Mechanics and its Applications}, 287(3):374--382, 2000.

\bibitem{preis2012quantifying}
Tobias Preis, Dror~Y Kenett, H~Eugene Stanley, Dirk Helbing, and Eshel Ben-Jacob.
\newblock Quantifying the behavior of stock correlations under market stress.
\newblock {\em Scientific reports}, 2(1):752, 2012.

\bibitem{SILVA2004227}
A.~Christian Silva, Richard~E. Prange, and Victor~M. Yakovenko.
\newblock Exponential distribution of financial returns at mesoscopic time lags: a new stylized fact.
\newblock {\em Physica A: Statistical Mechanics and its Applications}, 344(1):227--235, 2004.
\newblock Applications of Physics in Financial Analysis 4 (APFA4).

\bibitem{2024arXiv240211930T}
Yaoyue {Tang}, Karina {Arias-Calluari}, Michael~S. {Harr{\'e}}, and Fernando {Alonso-Marroquin}.
\newblock {Stylized Facts of High-Frequency Bitcoin Time Series}.
\newblock {\em arXiv e-prints}, page arXiv:2402.11930, February 2024.

\bibitem{tassywhite}
Martin {Tassy} and David {White}.
\newblock {Growth rate of a liquidity provider's wealth in $xy=c$ automated market makers}, 2020.

\bibitem{seang2019proof}
Dominique Torre and Sothearath Seang.
\newblock {Proof of Work and Proof of Stake consensus protocols: a blockchain application for local complementary currencies}.
\newblock working paper or preprint, August 2019.

\bibitem{ZHANG2019598}
Yuanyuan Zhang, Stephen Chan, Jeffrey Chu, and Saralees Nadarajah.
\newblock Stylised facts for high frequency cryptocurrency data.
\newblock {\em Physica A: Statistical Mechanics and its Applications}, 513:598--612, 2019.

\bibitem{Z2009}
Gilles Zumbach.
\newblock Time reversal invariance in finance.
\newblock {\em Quantitative Finance}, 9(5):505--515, 2009.

\end{thebibliography}
\end{document}